\documentclass{article}
\usepackage[utf8]{inputenc}
\usepackage{authblk}
\usepackage{times}
\usepackage{setspace}
\usepackage[margin=1.25in]{geometry}
\usepackage{graphicx}
\graphicspath{{fig/}}
\usepackage{subcaption}
\usepackage{amsmath}
\usepackage{float}
\usepackage{array}
\usepackage{url}

\usepackage[style=nejm,
citestyle=numeric-comp,
sorting=none,
doi=false,
url=false,
isbn=false,
eprint=false]{biblatex}
\AtBeginBibliography{\sloppy}

\newcounter{subpanel}
\newcommand{\resetpanels}{\setcounter{subpanel}{0}}
\newcommand{\subpanel}[2]{%
    \stepcounter{subpanel}%
    \begin{minipage}{#1}%
        \centering
        \includegraphics[width=\linewidth]{#2}\\[-0.25em]
        {\footnotesize (\alph{subpanel})}%
    \end{minipage}%
}

\title{Full-Path Nonlinear Modeling of Microwave Power Transmission Through Ionospheric Plasma for Space Solar Power Station}

\author[1,3]{Pengan Guo}
\author[1,2*]{Lei Chang}
\author[1]{Yuhan Chen}
\author[1]{Ya Gao}
\author[1]{Longshuai Ye}
\author[1]{Jikai Sun}
\author[1,2*]{Huaiqing Zhang}
\author[1,2]{Jian Li}

\affil[1]{School of Electrical Engineering, Chongqing University, Chongqing 400044, China.}
\affil[2]{Institute of Nuclear Energy and Technology Innovation, Chongqing University, Chongqing 400044, China.}
\affil[3]{Hongshen Honors School, Chongqing University, Chongqing 400044, China.}
\affil[*]{Correspondence emails: leichang@cqu.edu.cn; zhanghuaiqing@cqu.edu.cn.}

\date{}

\begin{document}

\nocite{Glaser1968Power,DOENASA1978SPS,Mankins2012SPSAlpha,Brown1984RadioPower,Shinohara2007Wireless,Shinohara2013Rectennas,Sasaki2013MPT,Malaviya2022SSPSReview,Budden1985,Schunk2009Ionospheres,Gurevich1978,Hasegawa1975Plasma,Akhiezer1975Plasma,Duncan1977IonosphereMicrowave,Perkins1978Ionospheric,Matsumoto1982SPS,Kaya1981SpaceChamber,Kaya1986MINIX,Nagatomo1986MINIX,Matsumoto1986Cyclotron,Matsumoto1995Nonlinear,Perkins1974Parametric,Thome1974Striations,Perkins1981SelfFocusing,Shinohara1995SelfFocusing,Nakamoto2007Cavitation,Omura2005SSPS,Monk2003MaxwellFEM,Jin2014FEMElectromagnetics,Berenger1994PML,Picone2002,Emmert2021NRLMSIS20,Banks1973Aeronomy,Raissi2019PINN,Bonzanini2023MLLTP,Zhong2022LTPPINN,Bilitza2017,Bilitza2022IRIReview,Fritsch1980PCHIP}

\maketitle

\begin{abstract}
Space Solar Power Station (SSPS) concepts rely on gigawatt-class microwave beams to carry orbital solar energy through the ionosphere, where the beam and the plasma form a coupled nonlinear system: the field heats electrons, the heating alters the collision frequency and plasma density, and the modified medium in turn reshapes the field. To our knowledge, this work is the first study to quantify this two-way interaction between microwave power transmission and the ionospheric plasma environment through full-path nonlinear modeling. The 340 km path from 400 km to 60 km altitude is reconstructed by 34 cascaded two-dimensional axisymmetric finite-element full-wave segments with complex-field transfer, using International Reference Ionosphere (IRI) electron-density and NRLMSISE-00 neutral-atmosphere inputs. A Shallow Neural Network (SNN) surrogate replaces the implicit electron energy balance with an explicit closure that maps altitude and local field magnitude to electron temperature and effective collision frequency, enabling stable nonlinear iteration. For 1 GW beams at 2.45 GHz and 5.8 GHz, the volume-integrated Ohmic deposition is 29.4 kW and 5.11 kW, respectively---fractional losses of order $10^{-5}$---and the ratio between the two bands follows the $\omega^{-2}$ scaling of collisional absorption. The deposition concentrates near 95 km altitude, where the product of electron density and collision frequency peaks, whereas the electron-temperature perturbation (up to 3815 K) maximizes in the F region, where cooling is weakest; ponderomotive density depletion remains below 0.02\%. The ionosphere is therefore effectively transparent to the SSPS power budget but not to the beam phase: localized heating and refractive perturbation accumulate phase-front distortion relevant to phased-array beam control, rectenna phase compensation, and environmental assessment.
\end{abstract}



\section{Introduction}
A Space Solar Power Station (SSPS) replaces the transmission line of a conventional power plant with a microwave beam. Since Glaser first proposed collecting solar energy in orbit to bypass the day--night cycle, weather, and land-use limits of ground solar farms \cite{Glaser1968Power}, and the DOE/NASA Satellite Power System studies developed the concept into an end-to-end energy system \cite{DOENASA1978SPS}, the microwave beam has been understood not as a communication carrier but as the principal energy-transport channel between an orbital power plant and a ground rectenna. Modern modular architectures such as SPS-ALPHA \cite{Mankins2012SPSAlpha}, together with decades of progress in microwave power transmission (MPT), solid-state sources, digital phased arrays, and rectenna technology \cite{Brown1984RadioPower,Shinohara2007Wireless,Shinohara2013Rectennas,Sasaki2013MPT,Malaviya2022SSPSReview}, have made system-level SSPS studies credible. Within this design space, frequencies near 2.45 GHz and 5.8 GHz remain the leading candidates, balancing antenna aperture, beam divergence, atmospheric transmission, component maturity, and rectenna efficiency.

This power line, however, runs through a plasma. For ordinary communication links, the ionosphere is adequately treated by weak linear phase and absorption corrections \cite{Budden1985,Schunk2009Ionospheres}. At gigawatt power levels the same medium becomes active and nonlinear: collisional Ohmic heating raises the electron temperature, the heated electrons change the collision frequency and hence the complex permittivity, the ponderomotive force redistributes the plasma density, and the perturbed refractive index feeds back on the beam itself, with possible coupling to electrostatic modes and parametric instabilities \cite{Gurevich1978,Hasegawa1975Plasma,Akhiezer1975Plasma,Perkins1974Parametric}. For SSPS, the ionosphere is therefore not a fixed attenuation layer but a nonlinear dielectric that responds to the very beam it transmits.

This concern has accompanied SSPS studies for decades, and the essential mechanisms are individually well documented. Early beam-interaction assessments identified the power densities at which nonlinear ionospheric responses become observable \cite{Duncan1977IonosphereMicrowave}; Perkins and Roble predicted substantial radio-wave heating below about 200 km \cite{Perkins1978Ionospheric}, and Matsumoto estimated the ionospheric impact of SPS microwaves numerically \cite{Matsumoto1982SPS}. Space-chamber experiments and the MINIX rocket campaign demonstrated that intense microwave beams couple energy into Langmuir and electron-cyclotron waves and produce local Ohmic heating \cite{Kaya1981SpaceChamber,Kaya1986MINIX,Nagatomo1986MINIX,Matsumoto1986Cyclotron,Matsumoto1995Nonlinear}, motivating studies of parametric instability and three-wave coupling \cite{Hasegawa1975Plasma,Akhiezer1975Plasma,Matsumoto1995Nonlinear,Perkins1974Parametric} as well as of self-focusing and ponderomotive filamentation \cite{Gurevich1978,Thome1974Striations,Perkins1981SelfFocusing,Shinohara1995SelfFocusing,Nakamoto2007Cavitation}. Threshold analyses found that a reference SPS beam at 2.45 GHz, with center field near 200 V/m, lies well below the strong self-focusing regime \cite{Thome1974Striations,Shinohara1995SelfFocusing}, and particle-in-cell environmental studies indicated that localized electrostatic activity and electron heating can coexist with 90--100\% pump-wave transmission in scaled cases \cite{Omura2005SSPS}. Yet each of these studies isolates a local mechanism, threshold, or environmental consequence under reduced geometry, scaled frequency ratios, or idealized plasma assumptions.

What remains unanswered is the propagation-scale question at the heart of SSPS beam transport: how much does the real, altitude-structured ionosphere affect a gigawatt-class power beam over its full path---and how much does the beam affect the ionosphere in return---when diffraction, phase accumulation, thermal feedback, and field-driven density perturbation act together? A direct full-wave answer is numerically prohibitive: the 340 km path spans millions of carrier wavelengths, and the problem is nested---the field sets the Ohmic heating rate, the heating sets the electron temperature, the temperature sets the collision frequency, and the updated collision frequency and density enter the Drude--Lorentz permittivity that governs the next field solution.

This paper closes that gap. To the best of our knowledge, it is the first study to quantify the mutual effect between microwave power transmission and the ionospheric plasma environment through full-path nonlinear modeling. The physical path from 400 km to 60 km altitude is decomposed into 34 connected 10 km two-dimensional axisymmetric full-wave segments with complex-field transfer, so that amplitude and phase information are preserved along the entire path; realistic International Reference Ionosphere (IRI) electron-density and NRLMSISE-00 neutral-atmosphere profiles define the background medium. To make the nonlinear thermal closure practical inside this cascaded finite-element full-wave model (CFEM), the implicit electron energy balance is solved offline and represented by a Shallow Neural Network (SNN) surrogate, $T_e=f_\theta(z,|\mathbf{E}|)$, with a companion surrogate for the temperature-dependent collision frequency; the SNN does not replace Maxwell's equations or the plasma model but provides a fast explicit closure for the local thermal--collisional response. The framework is applied to 1 GW beams at 2.45 GHz and 5.8 GHz under the same background ionosphere, yielding the volume-integrated heat deposition, the electron-temperature and density perturbations, and the phase-relevant beam evolution over the full path, and thereby separating general ionospheric response channels from band-dependent behavior.

The remainder of this paper is organized as follows. Section~\ref{sec:model_methods} details the nonlinear governing equations, geometric setup, segmented propagation strategy, surrogate modeling approach, and auxiliary consistency analysis. Section~\ref{sec:results_discussion} analyzes the simulation results, including cross-model consistency, altitude-block Ohmic-heating diagnostics, beam-transfer behavior, and the physical interpretation and frequency scaling of the computed nonlinear responses. Section~\ref{sec:conclusion} summarizes the implications for high-power microwave transmission and nonlinear electromagnetic simulation.

\section{Governing Equations and Methods}
\label{sec:model_methods}
\subsection{Frequency-Domain Maxwell Solver and Beam-Envelope Formulation}
The full-wave calculation starts from Maxwell's equations in a source-free, weakly ionized medium. In the time domain, the governing equations are
\begin{subequations}
\begin{align}
\nabla\times\mathbf{E} &=-\frac{\partial\mathbf{B}}{\partial t},\\
\nabla\times\mathbf{H} &=\frac{\partial\mathbf{D}}{\partial t}+\mathbf{J},\\
\nabla\cdot\mathbf{D} &=0,\\
\nabla\cdot\mathbf{B} &=0,
\end{align}
\end{subequations}
where $\mathbf{E}$ and $\mathbf{H}$ are the electric and magnetic fields, $\mathbf{D}$ and $\mathbf{B}$ are the electric and magnetic flux densities, and $\mathbf{J}$ is the conduction current associated with collisional plasma loss. The medium is described by
\begin{equation}
\mathbf{D}=\epsilon_0\epsilon_r\mathbf{E},\qquad
\mathbf{B}=\mu_0\mu_r\mathbf{H},\qquad
\mathbf{J}=\sigma\mathbf{E},
\end{equation}
where $\epsilon_r$ and $\sigma$ are generally complex, altitude dependent, and field dependent through the nonlinear plasma closure. The ionospheric plasma is nonmagnetic in the present frequency range, so $\mu_r=1$ is used in the numerical calculation.

For a monochromatic microwave beam, all field quantities are written as complex phasors with the time convention $\exp(i\omega t)$. Maxwell's curl equations then become
\begin{subequations}
\begin{align}
\nabla\times\mathbf{E} &=-i\omega\mu_0\mu_r\mathbf{H},\\
\nabla\times\mathbf{H} &=\left(i\omega\epsilon_0\epsilon_r+\sigma\right)\mathbf{E}.
\end{align}
\end{subequations}
Eliminating $\mathbf{H}$ gives a single vector wave equation for the electric field. For each propagation block, the unknown microwave field satisfies the curl--curl form
\begin{equation}
\nabla\times\left(\mu_r^{-1}\nabla\times\mathbf{E}\right)
-k_0^2\left(\epsilon_r-\frac{i\sigma}{\omega\epsilon_0}\right)\mathbf{E}=0,
\end{equation}
where $k_0=\omega/c$ is the free-space wavenumber. The term in parentheses may be regarded as an effective complex relative permittivity,
\begin{equation}
\epsilon_{\mathrm{eff}}=\epsilon_r-\frac{i\sigma}{\omega\epsilon_0},
\end{equation}
which represents the effective material coefficient in a general conductive-medium formulation. In the ionospheric-plasma calculation below, this coefficient is evaluated through the collision-dependent Drude--Lorentz response, so collisional loss is embedded in the plasma permittivity rather than added as an independent empirical absorption term. The effective material coefficient is recalculated from the local electron density, collision frequency, electron temperature, and field-driven density perturbation, so the Maxwell problem and the plasma closure form a nonlinear electromagnetic-material system.

The physical propagation distance is much longer than the microwave wavelength. A 2.45 GHz wave has a free-space wavelength of about 12 cm and a 5.8 GHz wave about 5.2 cm, while the modeled ionospheric path spans 340 km from 400 km to 60 km altitude. In the CFEM model, this full path is divided into 10 km propagation segments. Directly resolving every carrier oscillation over the full path would lead to an excessive number of elements even before the nonlinear plasma feedback is considered. To reduce this scale separation while retaining the frequency-domain Maxwell description, the electric field is factorized into a rapidly varying reference phase and a slowly varying complex envelope,
\begin{equation}
\mathbf{E}(\mathbf{r})=\mathbf{E}_1(\mathbf{r})\exp[-i\phi(\mathbf{r})],
\end{equation}
where $\phi(\mathbf{r})$ is a prescribed eikonal phase associated with the nominal propagation direction, and $\mathbf{E}_1$ is the slowly varying complex envelope. In a straight propagation block, $\nabla\phi$ is chosen to follow the dominant axial propagation direction, so the known rapid phase accumulation is removed from the numerical unknown. Substitution of this representation into the curl--curl equation gives the beam-envelope equation
\begin{equation}
\left(\nabla-i\nabla\phi\right)\times
\mu_r^{-1}\left[\left(\nabla-i\nabla\phi\right)\times\mathbf{E}_1\right]
-k_0^2\left(\epsilon_r-\frac{i\sigma}{\omega\epsilon_0}\right)\mathbf{E}_1=0 .
\end{equation}
This equation has the same physical content as the original frequency-domain Maxwell equation, but the numerical unknown is now the envelope rather than the full rapidly oscillating carrier. The mesh is therefore required to resolve the radial beam variation, plasma-induced envelope modulation, diffraction, and boundary-layer absorption, rather than every carrier period along the full altitude block.

The envelope equation is discretized by the finite-element method in the 2D-axisymmetric $(r,z)$ domain \cite{Monk2003MaxwellFEM,Jin2014FEMElectromagnetics}. Multiplying the governing equation by a vector test function $\mathbf{v}$ and integrating over the computational domain $\Omega$ gives the weak form
\begin{equation}
\int_{\Omega}
\left[\left(\nabla-i\nabla\phi\right)\times\mathbf{v}\right]\cdot
\mu_r^{-1}\left[\left(\nabla-i\nabla\phi\right)\times\mathbf{E}_1\right]\,d\Omega
-\int_{\Omega} k_0^2\epsilon_{\mathrm{eff}}\,\mathbf{v}\cdot\mathbf{E}_1\,d\Omega
=0,
\end{equation}
with boundary terms treated through the port and absorbing-boundary conditions. Because the launched microwave is represented by the azimuthal electric component in the axisymmetric model, the primary unknown is the complex envelope $E_{\phi,1}(r,z)$. The physical electric field used in the plasma heating and material update is reconstructed as
\begin{equation}
\mathbf{E}(r,z)=\mathbf{E}_1(r,z)\exp[-i\phi(r,z)].
\end{equation}
The cycle-averaged electric-field magnitude $|\mathbf{E}|$ is then used to evaluate the Ohmic heating rate and the field-dependent ponderomotive density response.

The nonlinear solution is obtained by iterating between the Maxwell solve and the plasma material update. At a given iteration, the current values of $N_e$, $T_e$, and $\nu_{eff}$ define $\epsilon_{\mathrm{eff}}$ through the Drude--Lorentz response. The beam-envelope Maxwell equation is solved for $\mathbf{E}_1$, the physical field magnitude is reconstructed, and the local heating rate is evaluated. The surrogate thermal closure then returns an updated electron temperature and collision response, while the ponderomotive relation updates the field-dependent density perturbation. These updated plasma quantities are inserted back into the effective permittivity, and the Maxwell equation is solved again. The iteration continues until the envelope field and the material coefficients are mutually consistent within the nonlinear solver tolerance.

The incoming wave is imposed as a complex envelope on the entrance boundary. For the first block, this field is obtained from the Gaussian-tapered aperture model described below. For later blocks, it is defined by the complex field obtained at the end of the previous block. Absorbing layers are placed on the radial and exit boundaries so that outgoing components leave the local domain with minimal artificial reflection. After convergence, the exit-plane field is reconstructed with its phase factor and used to define the incident envelope for the next altitude block. In this way, the Maxwell solution preserves both amplitude and phase accumulation while making the long ionospheric path computationally tractable.

\subsection{Nonlinear Dielectric and Ponderomotive Models}
To characterize the interaction between the high-power microwave beam and the ionospheric plasma, two primary nonlinear mechanisms are incorporated into the self-consistent governing equations.

\subsubsection{Drude--Lorentz Dispersion Model}
The ionosphere is treated as a dispersive and lossy medium. The complex relative permittivity $\epsilon_r$ is described by the Drude--Lorentz model, which accounts for electron-temperature-dependent collisions \cite{Budden1985}:
\begin{equation}
\epsilon_r = 1 - \frac{\omega_p^2}{\omega(\omega - i\nu_{eff}(T_e))},
\end{equation}
where $\omega$ is the angular frequency of the microwave, and $\omega_p = \sqrt{N_e e^2/(\varepsilon_0 m_e)}$ is the plasma frequency. The effective collision frequency $\nu_{eff}$ is the sum of electron-neutral collisions $\nu_{en}$ and electron-ion Coulomb collisions $\nu_{ei}$. Crucially, $\nu_{eff}$ is a function of the local electron temperature $T_e$, which is solved via the energy balance equation under microwave heating, forming a nonlinear feedback loop between the electromagnetic field and the medium properties.
Thus, in the numerical plasma update, collisional absorption is represented by the imaginary part of the Drude--Lorentz response associated with $\nu_{eff}$; no separate empirical absorption term is added on top of this plasma response.

\subsubsection{Ponderomotive Force Effect}
At GW-level power densities, the spatial gradient of the microwave electric field exerts a ponderomotive force on the plasma, leading to a redistribution of the electron density. The ponderomotive potential $\Phi_p$ is expressed as \cite{Gurevich1978}:
\begin{equation}
\Phi_p = \frac{e^2 |\mathbf{E}|^2}{4m_e\omega^2}.
\end{equation}
Under the assumption of quasi-neutrality and steady-state pressure balance, the modified electron density $N_e$ is given by the Boltzmann distribution:
\begin{equation}
N_e = N_{e0}\exp\left(-\frac{\Phi_p}{k_B T_e}\right),
\end{equation}
where $N_{e0}$ is the background electron density and $k_B$ is the Boltzmann constant. This effect accounts for the density depletion in regions of high field intensity, which can induce additional phase distortion and self-focusing or defocusing effects as the beam propagates through the ionospheric plasma.

\subsection{Geometry}
\begin{figure}[H]
    \centering
    \includegraphics[width=\textwidth]{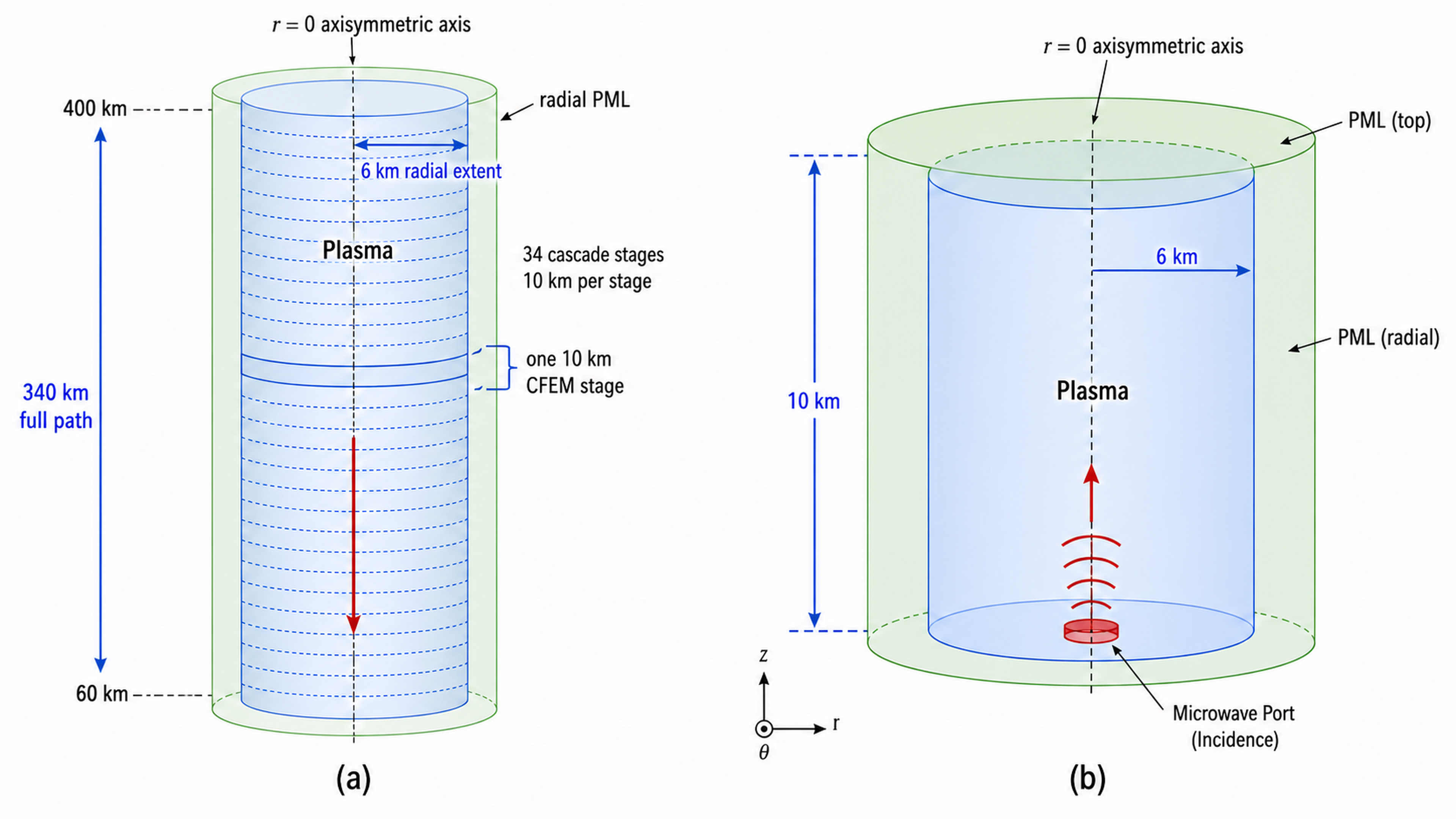}
    \caption{Schematic of the segmented full-path CFEM model: (a) full ionospheric cascade path and (b) local 10 km CFEM stage.}
    \label{fig:geometry}
\end{figure}

The CFEM geometry is organized as a segmented full-path model, as shown in Figure~\ref{fig:geometry}. The ionospheric transmission interval from 400 km to 60 km altitude is divided into 34 consecutive propagation segments, each representing a 10 km portion of the path. In each segment, the microwave beam is solved in a local 2D-axisymmetric finite-element domain with a 6 km radial extent, absorbing layers, and an incident-field boundary \cite{Berenger1994PML}. This segmented construction keeps each full-wave calculation computationally tractable while allowing the altitude-dependent plasma and neutral-atmosphere properties to be updated along the complete propagation path.

\subsection{Segmented Cascade Strategy for Long-Range Propagation}
As illustrated in Figure~\ref{fig:geometry}(a), resolving the full 340 km ionospheric path in one finite-element model would be computationally inefficient. The long-range propagation is therefore reconstructed by solving a sequence of local full-wave segments and connecting their complex beam fields. For the first segment, the incident field is generated from the 10 dB Gaussian-tapered aperture model with a 500 m aperture-edge radius and a total input power of 1 GW. For later segments, the complex field at the end of the preceding segment is mapped onto the entrance plane of the next segment, so that the radial beam profile and accumulated phase are retained rather than reset to an ideal Gaussian distribution.

In connecting adjacent segments, the propagated complex field is normalized so that each segment is evaluated at a reconstructed incident power of 1 GW. This choice reflects the physical regime considered here: the collisional thermal loss over each 10 km portion of the path is assumed to be negligible compared with the transmitted microwave power, and the results in Section~\ref{sec:results_discussion} confirm that the total Ohmic deposition remains small relative to the 1 GW beam. By contrast, using the cross-sectional power integral from each exit plane as the next input power can accumulate transfer errors associated with field interpolation, finite-domain representation, absorbing layers, and field projection between adjacent segments. These errors may exceed the small physical thermal loss that the model is designed to quantify. The normalized complex-field transfer therefore preserves the beam phase and radial structure for the nonlinear plasma update, while treating residual transfer loss as a diagnostic quantity rather than as the physical power delivered to the next segment.
\subsection{Electron Temperature and Surrogate Modeling}
The thermal response of the ionospheric plasma is determined by the local energy balance between Ohmic heating and cooling losses. The background atmospheric environment, including neutral species densities ($n_{N_2}$, $n_{O_2}$, $n_O$) and ambient temperature $T_n$, is obtained from the NRLMSISE-00 model \cite{Picone2002}, with geographic coordinates set to Bishan District, Chongqing. Recent updates of the MSIS empirical-atmosphere model family are described in \cite{Emmert2021NRLMSIS20}.

The Ohmic heating rate $Q_h$ represents the energy absorbed by electrons from the microwave field per unit volume, expressed as \cite{Perkins1978Ionospheric}:
\begin{equation}
Q_h = \frac{1}{2}\mathrm{Re}(\sigma)|\mathbf{E}|^2 = \frac{1}{2}\frac{N_e e^2\nu}{m_e(\omega^2+\nu^2)}|\mathbf{E}|^2,
\end{equation}
where $\sigma$ is the complex conductivity, $N_e$ is the electron density, $e$ and $m_e$ are the electron charge and mass, respectively, $\nu$ is the effective collision frequency, and $\omega$ is the microwave angular frequency. In the steady state, $Q_h$ is balanced by the total cooling rate $\sum_j L_{c,j}$ \cite{Banks1973Aeronomy}:
\begin{equation}
Q_h = \sum_j L_{c,j}(T_e,z,N_e).
\end{equation}
Under the cooling model adopted here, both the heating rate $Q_h$ and the included cooling terms $L_{c,j}$ scale with the local electron population. The temperature solution is therefore controlled primarily by altitude-dependent atmospheric composition, collision physics, and the local electric-field magnitude $|\mathbf{E}|$ over the sampled reference domain, rather than by an independently prescribed absolute density factor alone.

\begin{figure}[H]
    \centering
    \includegraphics[width=0.95\textwidth]{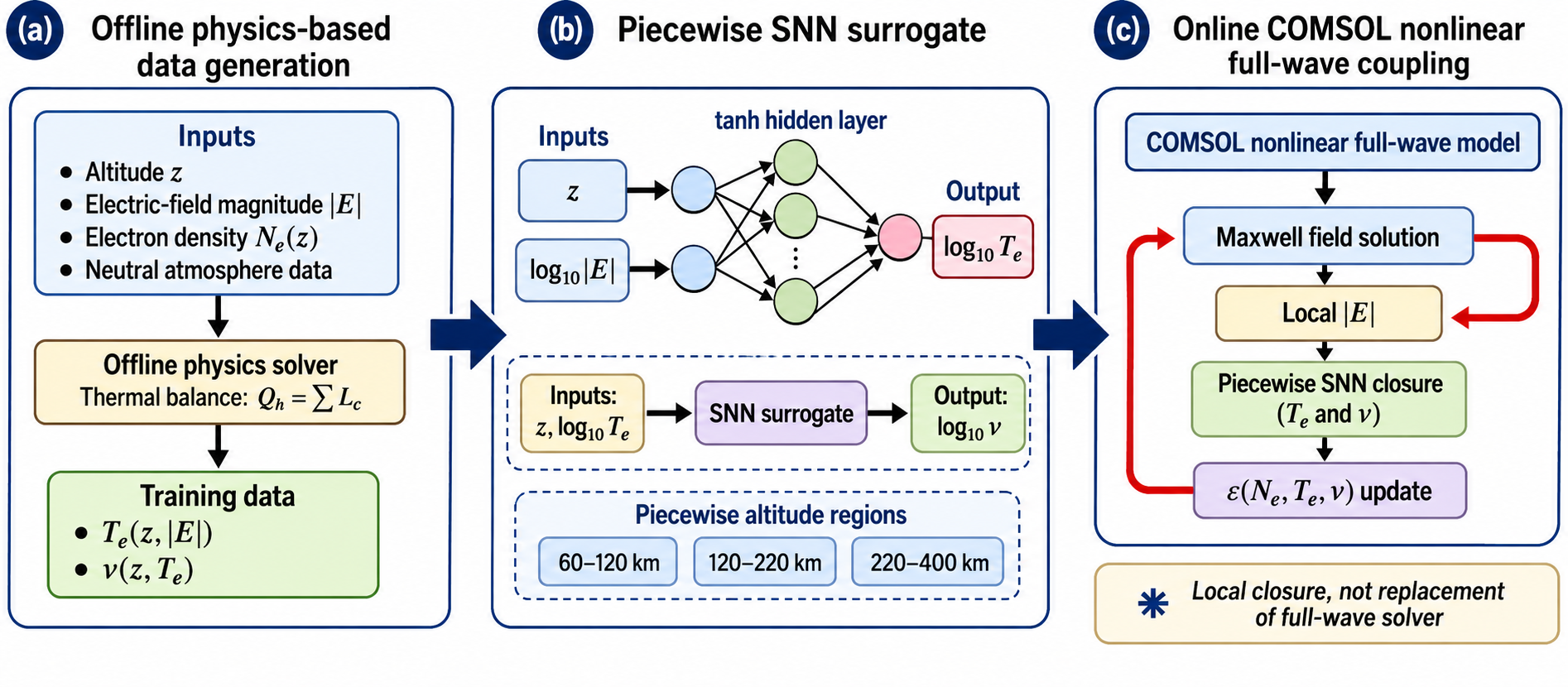}
    \caption{Workflow of the SNN surrogate closure used in the nonlinear full-wave model.}
    \label{fig:snn_workflow}
\end{figure}

To avoid the computational complexity of solving this implicit nonlinear equation directly within the nonlinear full-wave iterations, a surrogate modeling approach is adopted \cite{Raissi2019PINN,Bonzanini2023MLLTP,Zhong2022LTPPINN}. We first solve the balance equation numerically to generate a high-fidelity dataset of $T_e$ across a range of altitudes and field strengths. An SNN is then trained to fit this dataset, yielding an explicit and continuous mapping:
\begin{equation}
T_e = f(z, |\mathbf{E}|).
\end{equation}
This SNN-based explicit function is implemented as an analytic material closure in the CFEM solver, ensuring robust convergence and numerical stability during the self-consistent simulation of the nonlinear propagation process. The overall surrogate-closure structure is summarized in Figure~\ref{fig:snn_workflow}.

\subsubsection{SNN Surrogate Construction and Analytic Representation}
The surrogate is constructed offline before the full-wave calculation. The altitude domain is sampled from 60 km to 400 km with a 2 km interval, and the local electric-field magnitude is sampled logarithmically from $10^{-2}$ to 500 V/m. At each altitude--field pair, the nonlinear steady-state balance
\begin{equation}
F(T_e;z,|\mathbf{E}|)=Q_h(T_e,z,|\mathbf{E}|)-\sum_j L_{c,j}(T_e,z)=0
\end{equation}
is solved numerically. The lower bound of the search interval is the local neutral temperature, while the upper bound is selected adaptively from a sequence of candidate temperatures up to $10^5$ K. A bracketing root solver is then used whenever the heating and cooling terms change sign; otherwise the solution is assigned to the nearest physically admissible bound. This produces the reference surface $T_e(z,|\mathbf{E}|)$ used for training.

To improve numerical conditioning and avoid a single network having to represent the full altitude range at once, the surrogate is trained as a piecewise shallow neural network. The altitude interval is divided into three regions, 60--120 km, 120--220 km, and 220--400 km. For the electron-temperature surrogate, the corresponding hidden-layer widths are 18, 18, and 14 neurons. Each regional model is a single-hidden-layer multilayer perceptron with a hyperbolic-tangent activation function. The two inputs are altitude $z$ and $\log_{10}|\mathbf{E}|$, and the output is $\log_{10}T_e$. Both the inputs and output are standardized before training. The network weights are optimized using an L-BFGS solver with a small $L_2$ regularization coefficient, so the resulting function is smooth and deterministic.

The same procedure is used to construct the companion collision-frequency surrogate. In this case, the target surface is $\nu(z,T_e)$, the inputs are $z$ and $\log_{10}T_e$, and the output is $\log_{10}\nu$. The altitude regions are kept the same, with hidden-layer widths of 16, 16, and 12 neurons. The trained collision-frequency surrogate is then evaluated together with the electron-temperature surrogate inside the dielectric update, so that the Drude--Lorentz permittivity can respond explicitly to the local full-wave electric field.

\begin{figure}[H]
    \centering
    \resetpanels
    \subpanel{0.48\textwidth}{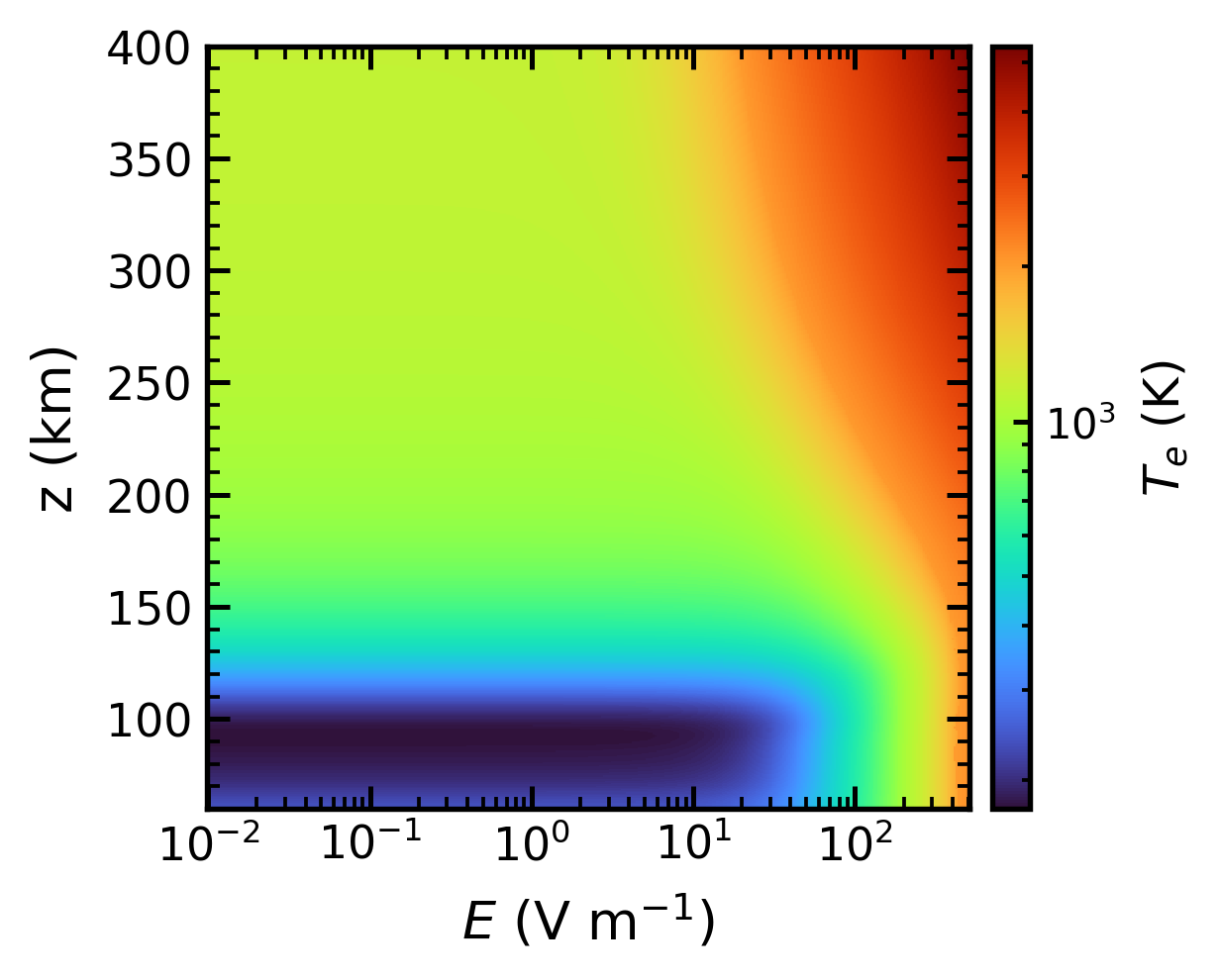}
    \hfill
    \subpanel{0.48\textwidth}{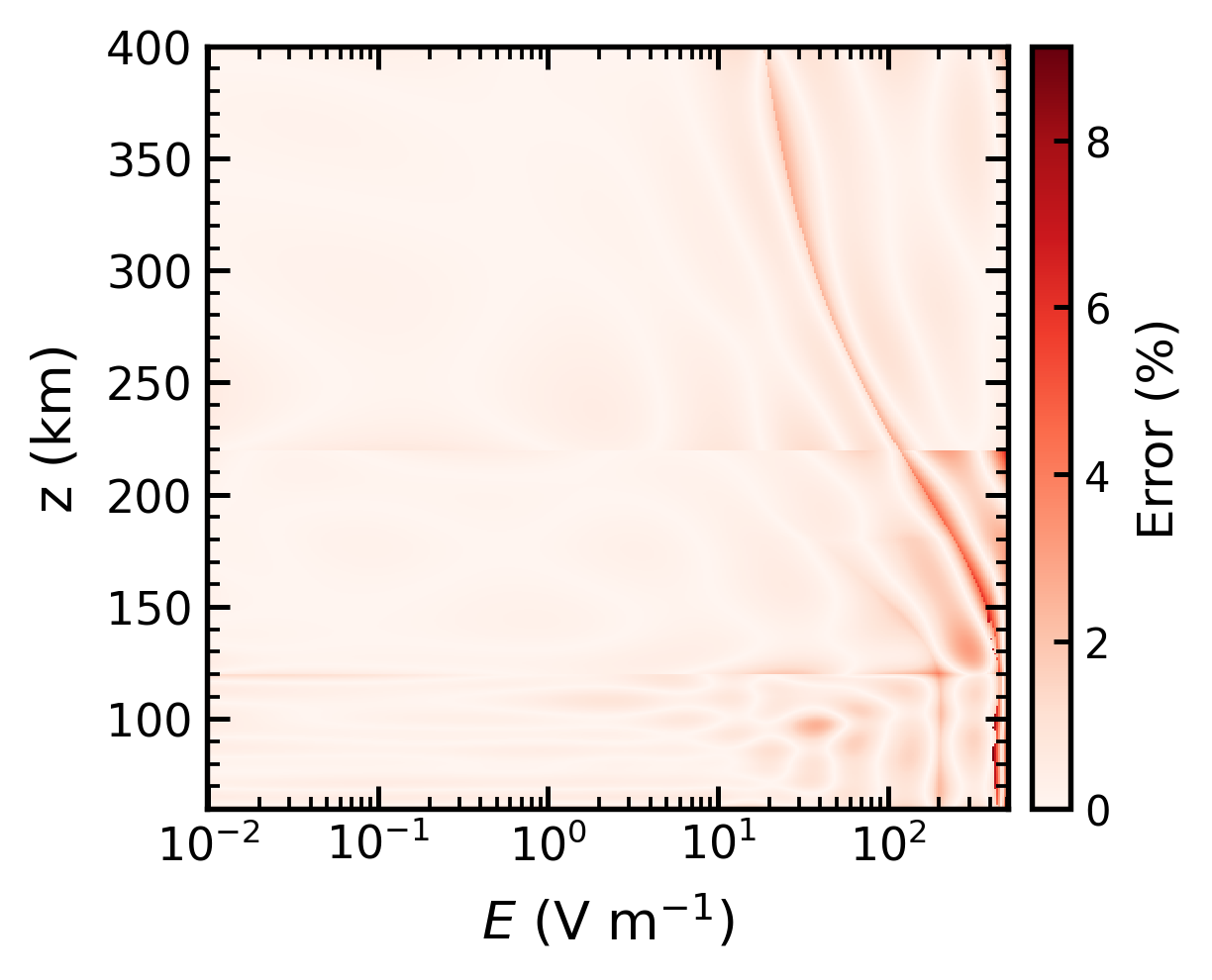}
    \caption{Performance of the 2.45 GHz SNN temperature surrogate: (a) electron temperature $T_e$ as a function of altitude and electric field, and (b) the corresponding neural-network fitting error.}
    \label{fig:snn_temperature_surface}
\end{figure}

After training, the fitted network is represented directly by closed-form analytic expressions during the nonlinear full-wave solve. The trained weights, biases, standardization constants, and piecewise altitude conditions define the local surrogate relation. For each altitude block, the expression has the form
\begin{equation}
\hat{y}=10^{a_0+\sum_j a_j\tanh(b_{j1}\tilde{x}_1+b_{j2}\tilde{x}_2+c_j)},
\end{equation}
where $\tilde{x}_1$ and $\tilde{x}_2$ are the standardized inputs, and $\hat{y}$ denotes either $T_e$ or $\nu$. The expressions are clipped to the training ranges to avoid extrapolation outside the sampled physical domain. This representation preserves the nonlinear thermal and collisional feedback while avoiding repeated local root solves during the finite-element nonlinear iterations. The fitted temperature surface and relative fitting error for the 2.45 GHz surrogate are shown in Figure~\ref{fig:snn_temperature_surface}.

\subsection{Electron Density Profile Reconstruction}
The electron density ($N_e$) is the most critical parameter governing the refractive index and the propagation characteristics of the ionospheric plasma. For this study, the background $N_e$ profiles are obtained from the IRI model through its Python interface, PyIRI \cite{Bilitza2017,Bilitza2022IRIReview}. The IRI model provides a standard empirical representation of ionospheric parameters based on geographic location, solar activity, and time.

\begin{figure}[H]
    \centering
    \resetpanels
    \subpanel{0.48\textwidth}{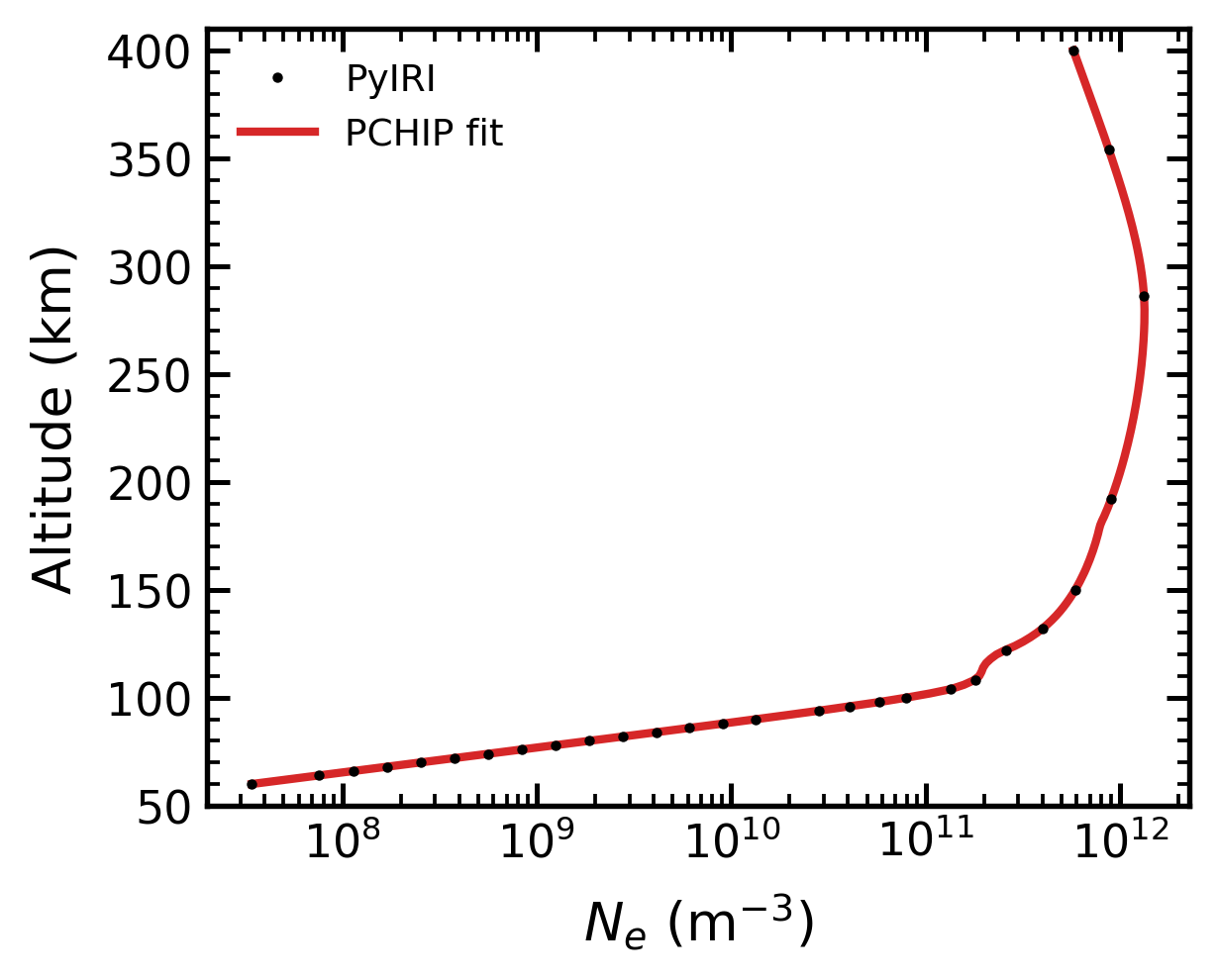}
    \hfill
    \subpanel{0.48\textwidth}{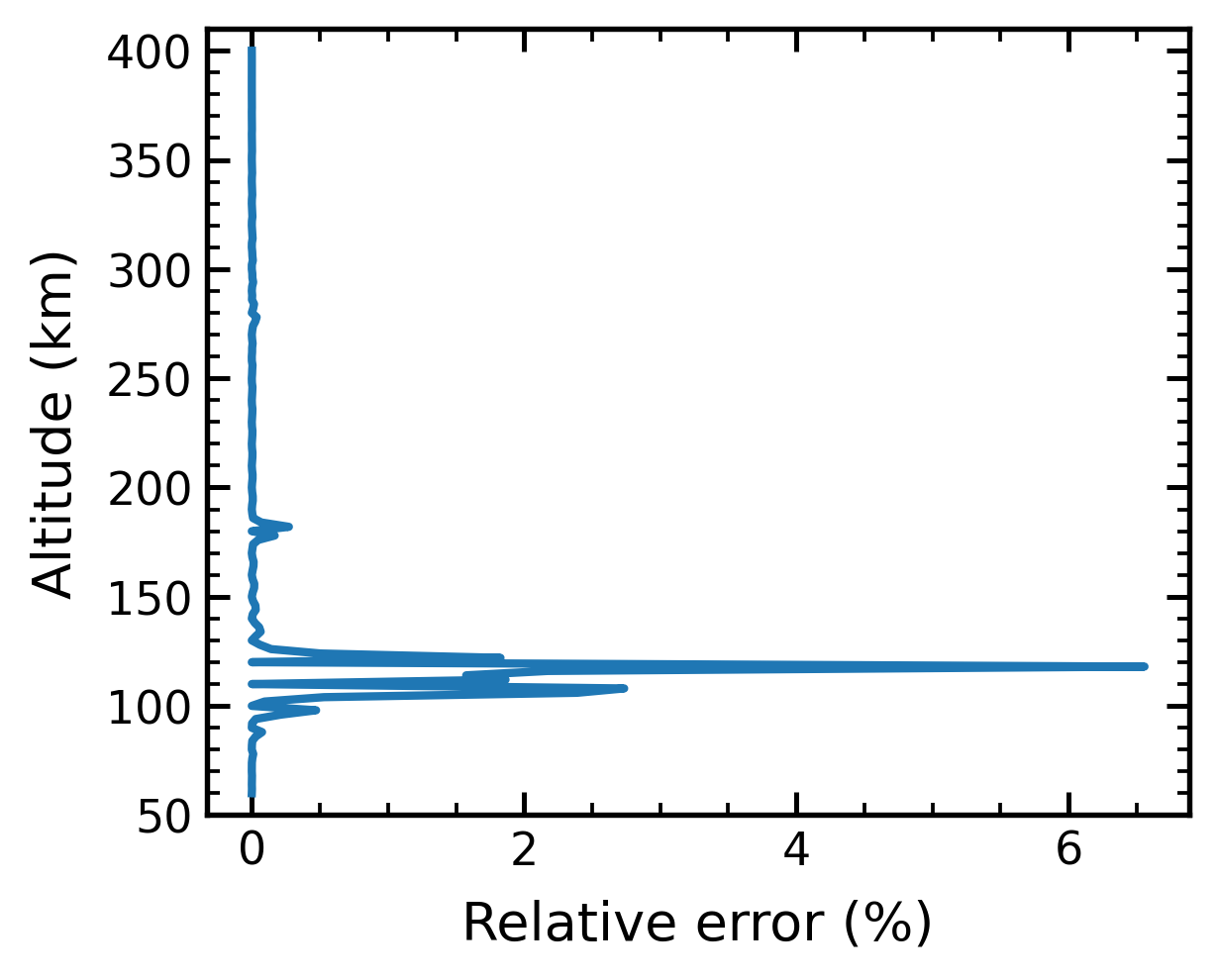}
    \caption{Ionospheric electron density profile derived from the IRI model and the numerical accuracy of the PCHIP interpolation scheme: (a) IRI profile and PCHIP fit, and (b) relative fitting error.}
    \label{fig:ElectronDensity}
\end{figure}

The vertical distribution of $N_e$ is characterized by a multi-layered structure, primarily featuring the E-layer and the F2-layer peak. To integrate the discrete data points from the PyIRI-derived IRI profile into the field solver, we employ a Piecewise Cubic Hermite Interpolating Polynomial (PCHIP) scheme \cite{Fritsch1980PCHIP}. The interpolation is performed in the logarithmic domain ($\log_{10}N_e$) to ensure positivity and to capture the sharp gradients present in the E-region without numerical oscillations. The local plasma frequency is derived as
\begin{equation}
\omega_p = \sqrt{\frac{N_e e^2}{\varepsilon_0 m_e}},
\end{equation}
where $e$ is the elementary charge, $\varepsilon_0$ is the vacuum permittivity, and $m_e$ is the electron mass. The fitted electron-density profile is shown in Figure~\ref{fig:ElectronDensity}.

\subsection{Collision Frequency}
The collision frequency $\nu$ is a fundamental parameter that dictates the rate of momentum and energy transfer between electrons and other particle species in the ionospheric plasma. The total effective collision frequency is generally expressed as the sum of electron-ion collisions and electron-neutral collisions:
\begin{equation}
\nu_{total} = \nu_{ei} + \sum_j \nu_{\mathrm{en},j}.
\end{equation}
Following Schunk and Nagy \cite{Schunk2009Ionospheres}, the electron-ion collision frequency is governed by Coulomb interactions:
\begin{equation}
\nu_{ei} = 54.5 \frac{n_i Z_i^2}{T_e^{3/2}}.
\end{equation}
Here $n_i$ is the ion number density and $Z_i$ is the ion charge state. For the singly ionized ionospheric plasma considered here, $Z_i=1$ is used unless otherwise stated.
For the lower and middle ionosphere, collisions with neutral species such as $\mathrm{N}_2$, $\mathrm{O}_2$, and $\mathrm{O}$ are dominant. The specific formulas employed are
\begin{subequations}
\label{eq:collision_frequencies}
\begin{align}
\nu_{\mathrm{en}}(\mathrm{N}_2) &= 2.33\times10^{-11} n(\mathrm{N}_2)(1-1.21\times10^{-4}T_e)T_e,\\
\nu_{\mathrm{en}}(\mathrm{O}_2) &= 1.82\times10^{-10} n(\mathrm{O}_2)(1+3.6\times10^{-2}T_e^{1/2})T_e^{1/2},\\
\nu_{\mathrm{en}}(\mathrm{O}) &= 8.9\times10^{-11} n(\mathrm{O})(1+5.7\times10^{-4}T_e)T_e^{1/2}.
\end{align}
\end{subequations}
where the neutral number densities in these empirical collision formulas are expressed in $\mathrm{cm}^{-3}$ and $T_e$ is in K. These empirical formulas are used within the temperature range relevant to the present ionospheric heating calculation, with the solved electron temperatures remaining on the order of $10^3$ K. All density-dependent quantities are converted to SI units before being coupled to the CFEM solver and the Ohmic-heating calculation.

\begin{figure}[H]
    \centering
    \resetpanels
    \subpanel{0.48\textwidth}{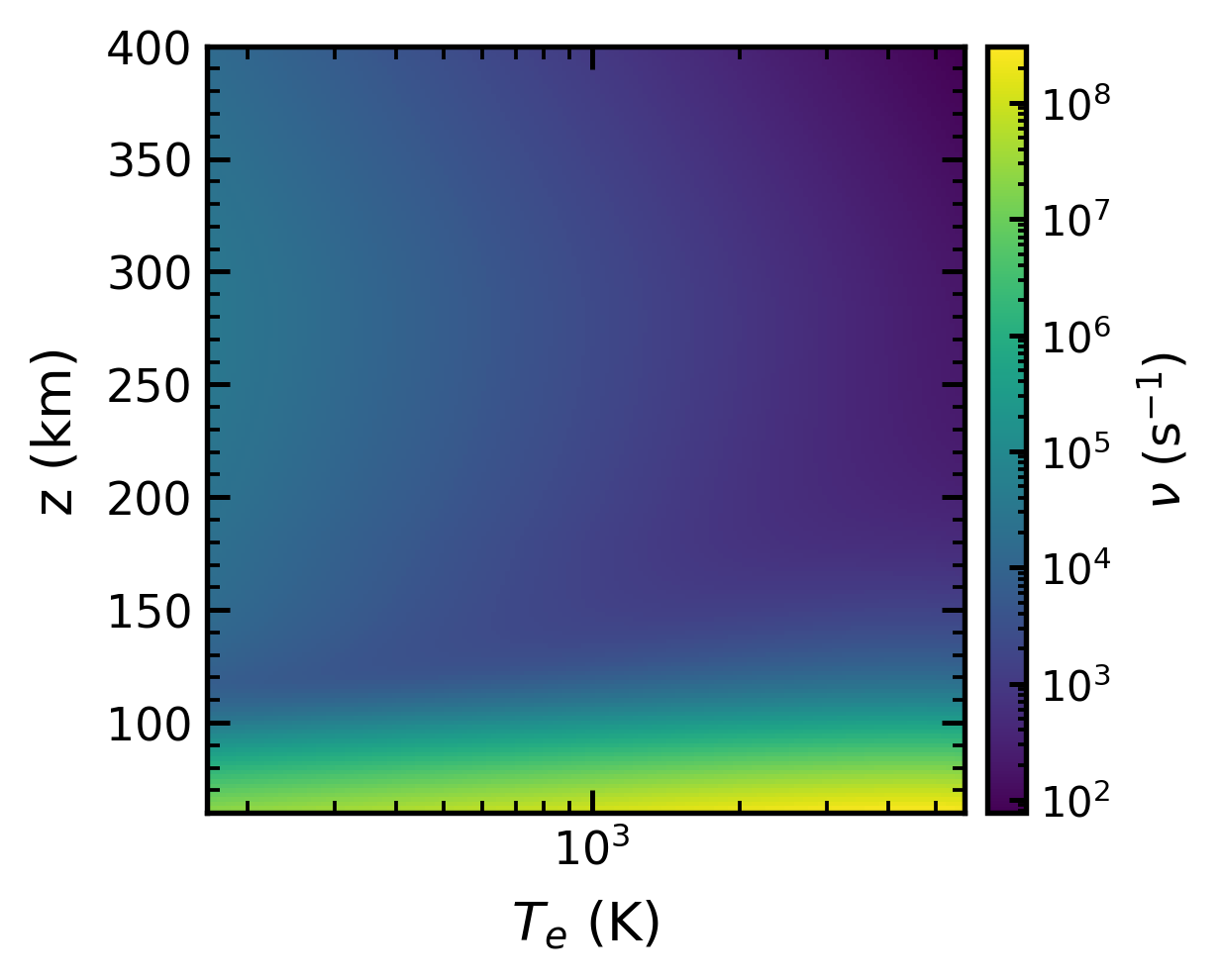}
    \hfill
    \subpanel{0.48\textwidth}{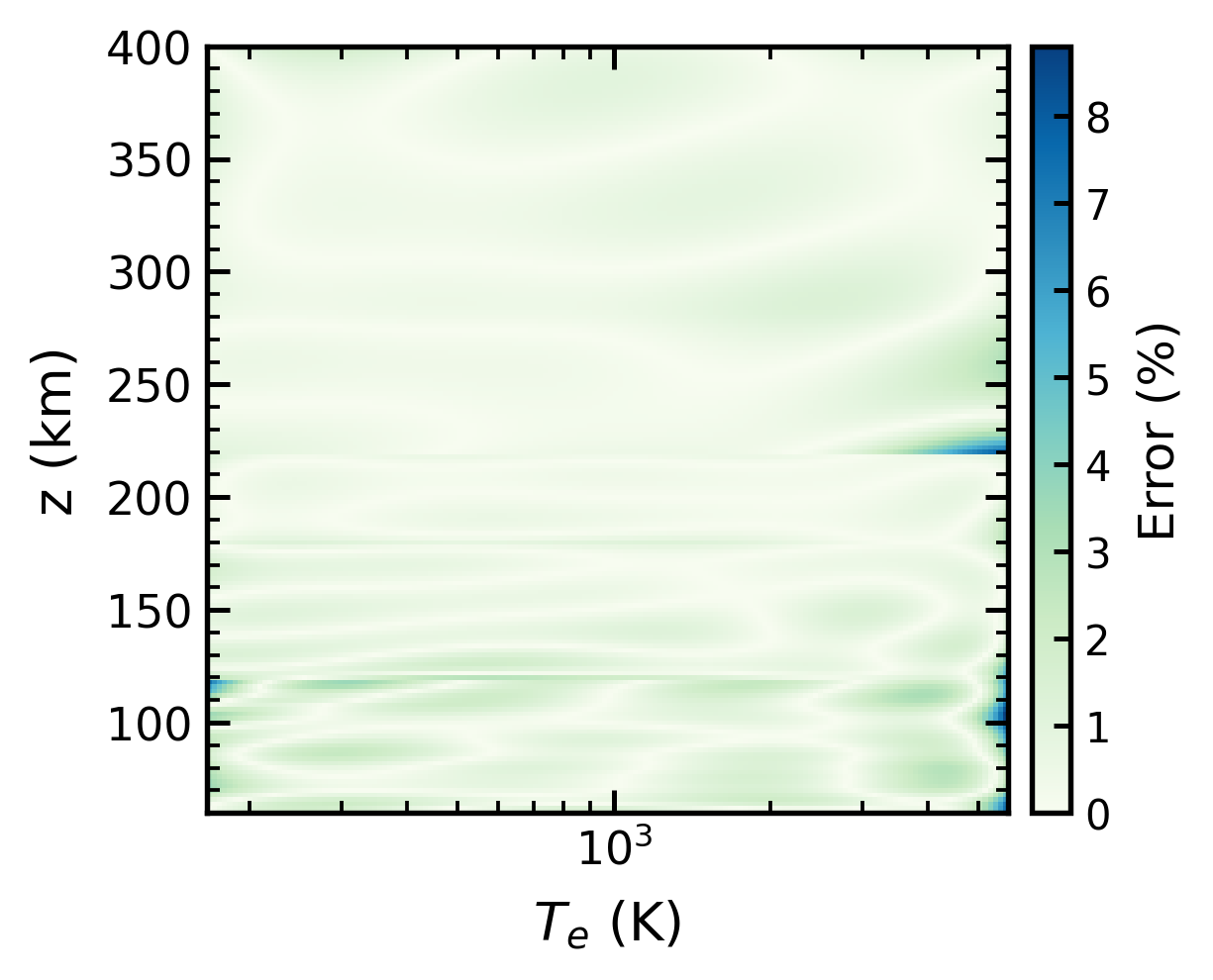}
    \caption{Collision-frequency surrogate data for the 2.45 GHz case: (a) physics-solver reference surface of the total collision frequency $\nu(z,T_e)$ as a function of altitude and electron temperature, and (b) relative fitting error.}
    \label{fig:nu_surface}
\end{figure}

The complexity and wide dynamic range of these equations pose significant challenges for robust convergence during finite-element nonlinear iterations. To address this, an SNN was trained using altitude $z$ and electron temperature $T_e$ as inputs to predict the local $\nu_{total}$. As illustrated in Figure~\ref{fig:nu_surface}, the 2.45 GHz collision-frequency surrogate surface spans several orders of magnitude.

Because the SNN is used as an analytic surrogate closure rather than as an extrapolative prediction model, its approximation accuracy is evaluated on the sampled reference grids used to construct the closed-form expressions. The error statistics in Table~\ref{tab:snn_accuracy} compare the closed-form SNN outputs with the corresponding physics-solver reference surfaces over the prescribed altitude--field and altitude--temperature domains.
\begin{table}[H]
\centering
\caption{SNN closure accuracy.}
\label{tab:snn_accuracy}
\renewcommand{\arraystretch}{1.15}
\setlength{\tabcolsep}{10pt}
\begin{tabular*}{0.92\textwidth}{@{\extracolsep{\fill}}lcccc@{}}
\hline
Frequency & $T_e$ mean (\%) & $T_e$ max. (\%) & $\nu$ mean (\%) & $\nu$ max. (\%) \\
\hline
2.45 GHz & 0.314 & 9.126 & 0.566 & 8.772 \\
5.8 GHz & 0.242 & 6.149 & 0.534 & 10.637 \\
\hline
\end{tabular*}
\end{table}

\subsection{Input Microwave Beam Profile}
The incident microwave beam is specified with a total input power $P_{\mathrm{in}}=1$ GW for the two operating frequencies considered in this work, $f=2.45$ GHz ($\lambda\approx0.1224$ m) and $f=5.8$ GHz ($\lambda\approx0.0517$ m). Following standard designs for space solar power transmission, the microwave beam at the transmitting aperture is modeled as a 10 dB Gaussian-tapered beam with a 500 m aperture-edge radius \cite{Shinohara2007Wireless}. The corresponding Gaussian beam radius evolves as
\begin{equation}
w(z)=w_0\sqrt{1+\left(\frac{z}{z_R}\right)^2},
\end{equation}
where $w_0$ is the Gaussian waist parameter implied by the selected aperture taper, and $z_R=\pi w_0^2/\lambda$ is the Rayleigh range. The electric field distribution at the input boundary is
\begin{equation}
E(r)=E_0\exp\left(-\frac{r^2}{w(z)^2}\right),
\end{equation}
and the peak electric field is
\begin{equation}
E_0=\sqrt{\frac{4\eta_0P_{\mathrm{in}}}{\pi w(z)^2}}.
\end{equation}
In the CFEM calculation, the incident field is normalized to the prescribed total input power after evaluating the Gaussian-tapered profile on the entrance radial grid. The 5.8 GHz case uses the same input power, aperture-edge radius, 10 dB Gaussian taper, propagation interval, and cascade-block settings as the 2.45 GHz case; only the carrier frequency, wavelength, and frequency-dependent Gaussian-beam and plasma-response quantities are updated. Because the shorter wavelength at 5.8 GHz increases the Rayleigh range for the same aperture setting, the vacuum-propagated beam radius at the 400 km entrance plane is smaller than in the 2.45 GHz case, so the on-axis Gaussian intensity is higher for the same transmitted power.

\subsection{Reduced-Order Propagation Diagnostic}
The primary electromagnetic solution in this work is obtained from the cascaded finite-element full-wave model (CFEM). An auxiliary Reduced-Order Propagation Diagnostic (ROPD) is used for input reconstruction, diagnostic calculation, and cross-model consistency checking. This separation is useful because the ROPD can rapidly evaluate one-dimensional altitude trends and analytic beam estimates, while the CFEM resolves the 2D-axisymmetric field distribution and nonlinear dielectric feedback. The distinction below refers to the numerical formulations: CFEM is the finite-element propagation route, whereas ROPD is an independent reference route.

The ROPD begins from the same physical parameter set used in the corresponding CFEM case: a 1 GW Gaussian-tapered microwave beam at either 2.45 GHz or 5.8 GHz, an altitude interval from 400 km to 60 km, and the same ionospheric and neutral-atmosphere background data. The electron-density profile is reconstructed on a uniform altitude grid from the PyIRI-derived profile and interpolated with a PCHIP scheme. The neutral temperature and the main neutral-species densities are evaluated consistently with the NRLMSISE-based thermal-balance model. From these profiles, the ROPD computes the local plasma frequency, propagation parameter $X=(\omega_p/\omega)^2$, effective collision frequency, complex permittivity, refractive index, and Ohmic heating rate.

For propagation diagnostics, the ROPD uses a reduced-order vertical Wentzel--Kramers--Brillouin (WKB) attenuation model,
\begin{equation}
\frac{dI}{dz}=-2k_0 n_{\mathrm{im}}(z)I,
\end{equation}
where $n_{\mathrm{im}}$ denotes the imaginary part of the complex refractive index. This notation is used here to avoid confusion with the ion number density $n_i$ in the collision-frequency model. This attenuation estimate is combined with the Gaussian-beam spreading model to provide the on-axis beam intensity along the ionospheric path. The electric-field magnitude inferred from the reduced-order intensity is then used in the same steady-state heating and cooling balance to estimate the electron-temperature perturbation. The ROPD also evaluates the ponderomotive density-depletion estimate and Gaussian radial beam profiles at selected altitudes. These quantities are not used to replace the CFEM solution; rather, they provide reference trends for interpreting the CFEM results and for assessing the consistency of the plasma inputs, beam normalization, and loss estimates.

The CFEM results are evaluated as centerline profiles, radial distributions, and two-dimensional field maps. The ROPD diagnostics are then compared with these quantities in the Results section. In this sense, the ROPD provides an auxiliary consistency-checking route around the full-wave model: it reconstructs the same background profiles, evaluates propagation and nonlinear-response indicators, and supplies the reference panels used in the cross-model comparisons. The role separation between the two routes is summarized in Table~\ref{tab:matlab_comsol_workflow}.

\begin{table}[H]
\centering
\caption{Role separation between the ROPD and CFEM routes.}
\label{tab:matlab_comsol_workflow}
\begin{tabular}{@{}l>{\raggedright\arraybackslash}p{0.39\textwidth}>{\raggedright\arraybackslash}p{0.39\textwidth}@{}}
\hline
Item & ROPD role & CFEM role \\
\hline
Profiles & Reconstruct $N_e(z)$, neutral species, and $T_n$ & Use interpolated profiles as material inputs \\
Propagation & Estimate on-axis intensity from Gaussian spreading and WKB absorption & Resolve 2D-axisymmetric full-wave field evolution \\
Plasma & Estimate diagnostic $Q_h$, $\Delta T_e$, and density-depletion trends & Update nonlinear dielectric properties self-consistently \\
Beam & Generate Gaussian radial reference profiles & Provide radial field and intensity profiles from the CFEM solution \\
Comparison & Provide reference trends & Provide CFEM quantities for comparison \\
\hline
\end{tabular}
\end{table}

\section{Results and Discussion}
\label{sec:results_discussion}
\subsection{For 2.45 GHz Propagation}
\subsubsection{Background Profiles and Cross-Model Consistency}
\begin{figure}[H]
    \centering
    \resetpanels
    \subpanel{0.32\textwidth}{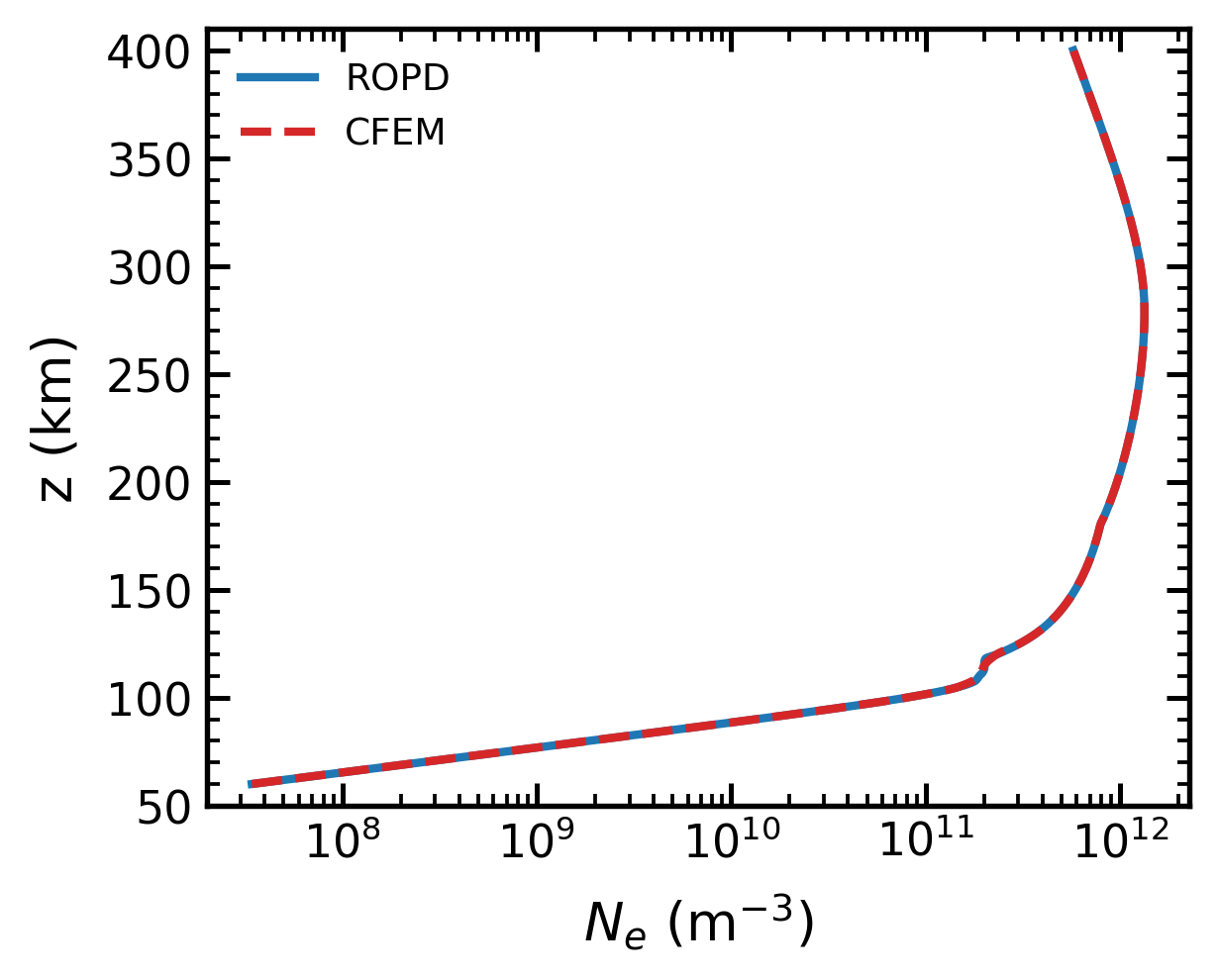}
    \hfill
    \subpanel{0.32\textwidth}{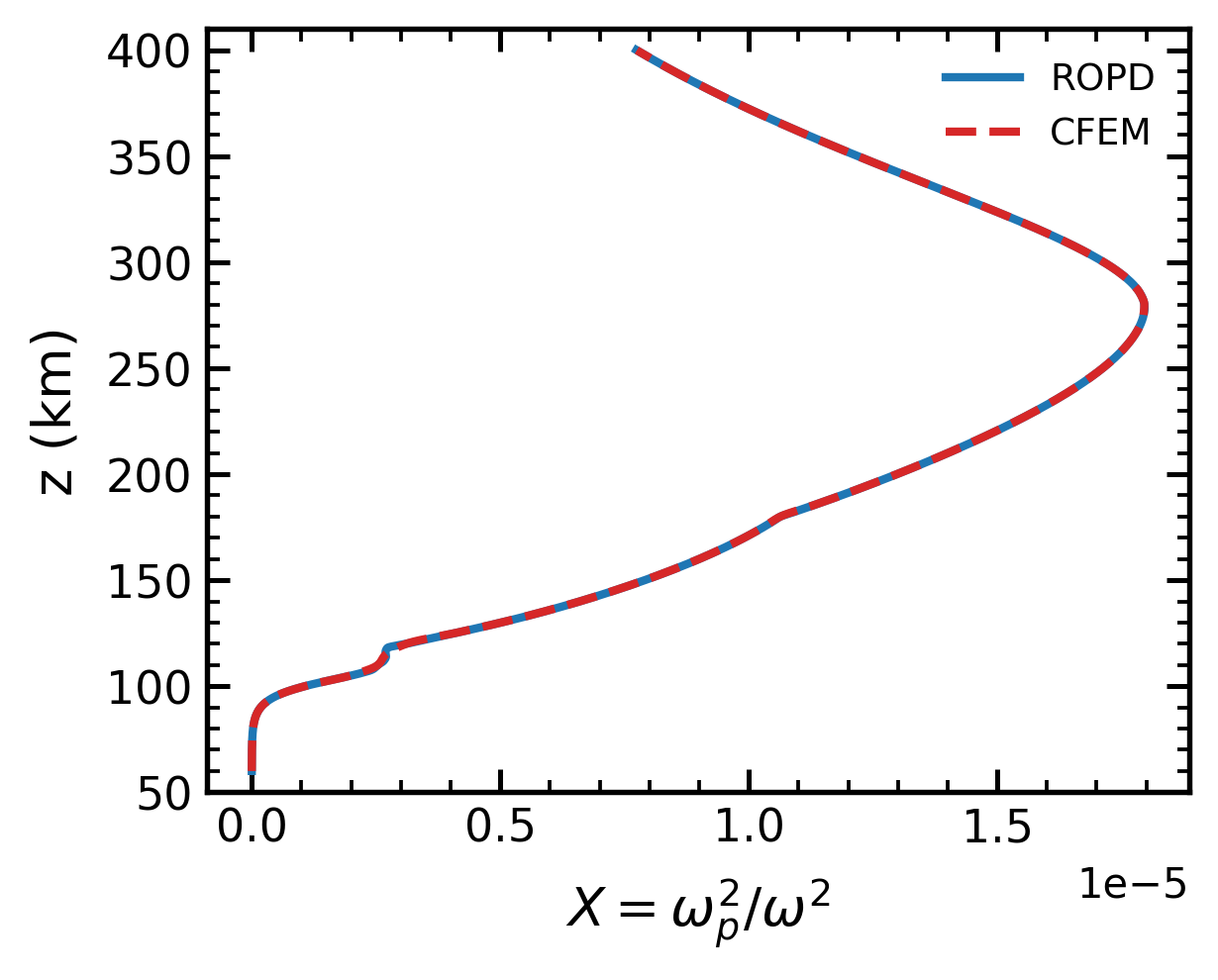}
    \hfill
    \subpanel{0.32\textwidth}{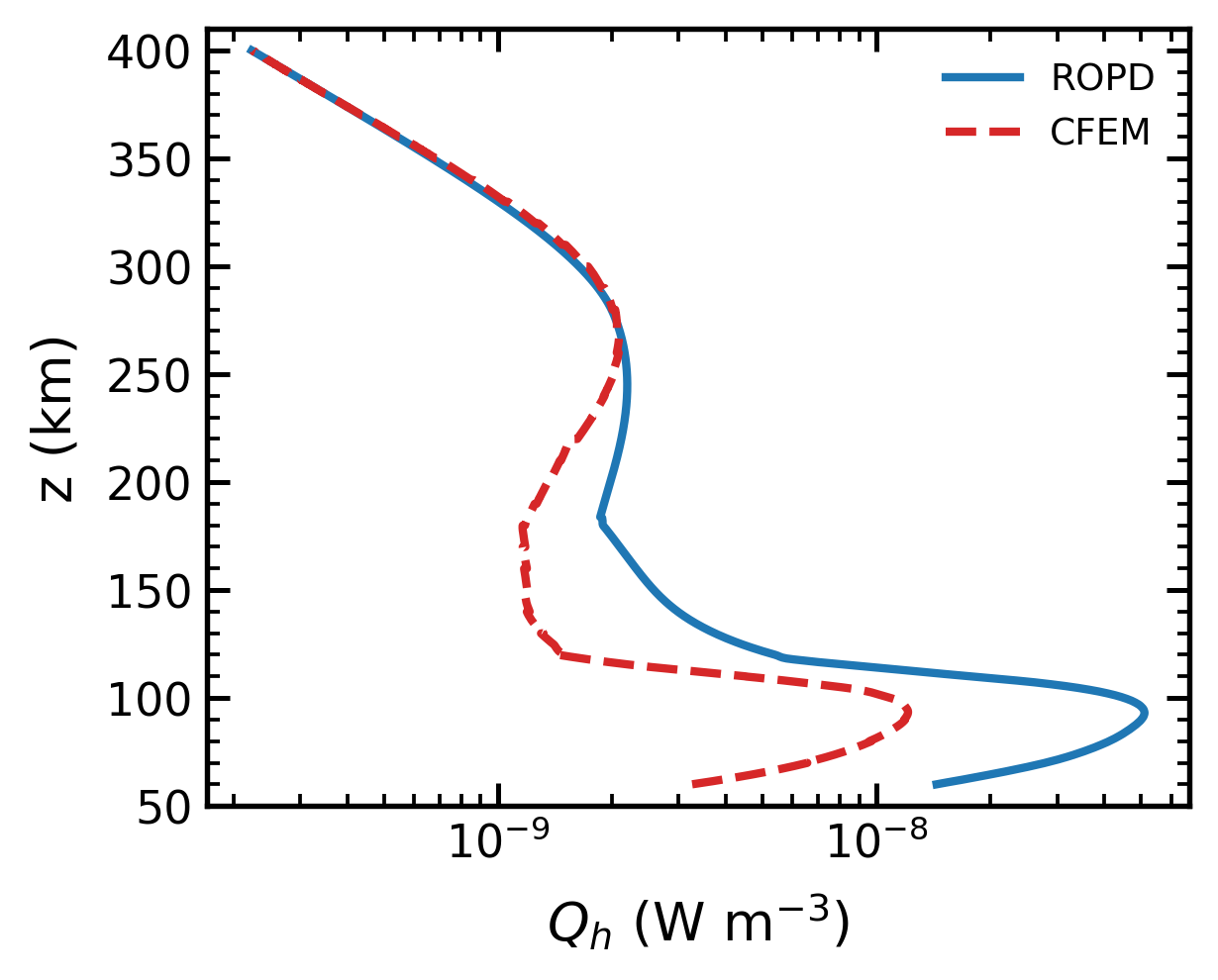}
    \caption{Background and loss-related profiles for the 2.45 GHz analysis: (a) electron density, (b) propagation parameter, and (c) Ohmic heating rate.}
    \label{fig:profile_matlab_comsol}
\end{figure}

Figure~\ref{fig:profile_matlab_comsol} summarizes the background and loss-related profiles used in the 2.45 GHz analysis. The electron density and propagation parameter provide common altitude-dependent inputs for the two routes, while the Ohmic heating rate illustrates how the reduced-order diagnostic and CFEM response begin to diverge once the local field distribution is included. The complex refractive index and electron temperature are calculated inside the CFEM route for the nonlinear dielectric update, while the beam intensity is discussed later as a model-dependent propagation response rather than as a direct shared input.

The propagation parameter $X=(\omega_p/\omega)^2$ remains far below unity along the entire path, so the beam is deeply underdense and never approaches a reflection layer; the interaction is perturbative in the dielectric sense even though the thermal response is strongly nonlinear. The altitude structure of the Ohmic heating rate follows directly from Equation~(13). Because $\nu\ll\omega$ holds everywhere along the modeled path at both carrier frequencies, the heating rate reduces to $Q_h\approx N_e e^2\nu|\mathbf{E}|^2/(2m_e\omega^2)$, so for a slowly varying beam intensity the deposition profile is controlled by the product $N_e(z)\,\nu(z)$. The collision frequency rises quasi-exponentially toward lower altitude with the neutral density, while the electron density collapses below the E region; the product is therefore sharply peaked in the lower E region near 90--100 km. In physical terms, the upper ionosphere contains electrons but almost no collisions to convert quiver energy into heat, and the lowest path segment contains collisions but almost no electrons: the lower E region is the only altitude band where both ingredients coexist, and it acts as a thin collisional absorption layer for the power beam.

\subsubsection{Nonlinear Heating and Plasma Response}
\begin{figure}[H]
    \centering
    \resetpanels
    \subpanel{0.48\textwidth}{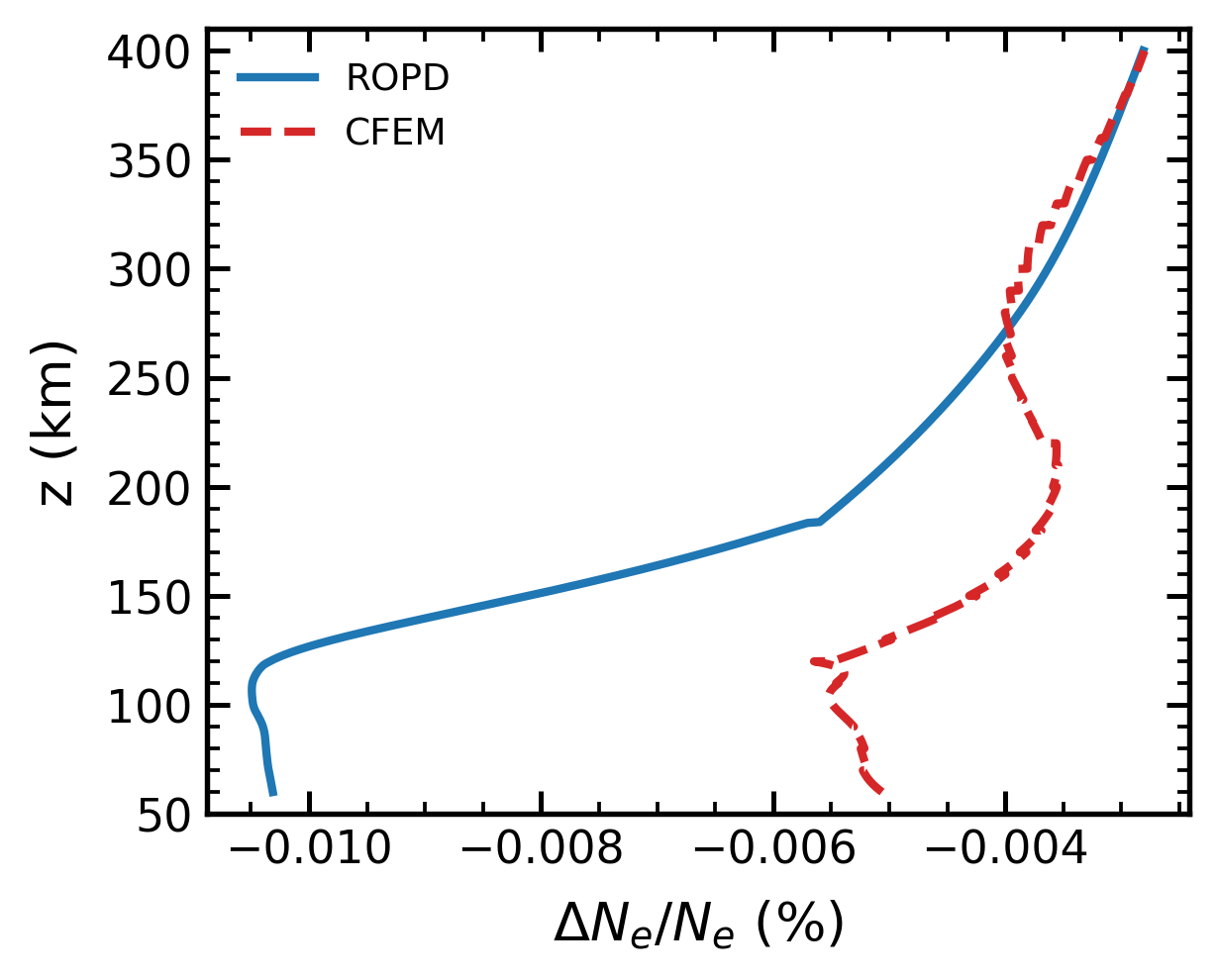}
    \hfill
    \subpanel{0.48\textwidth}{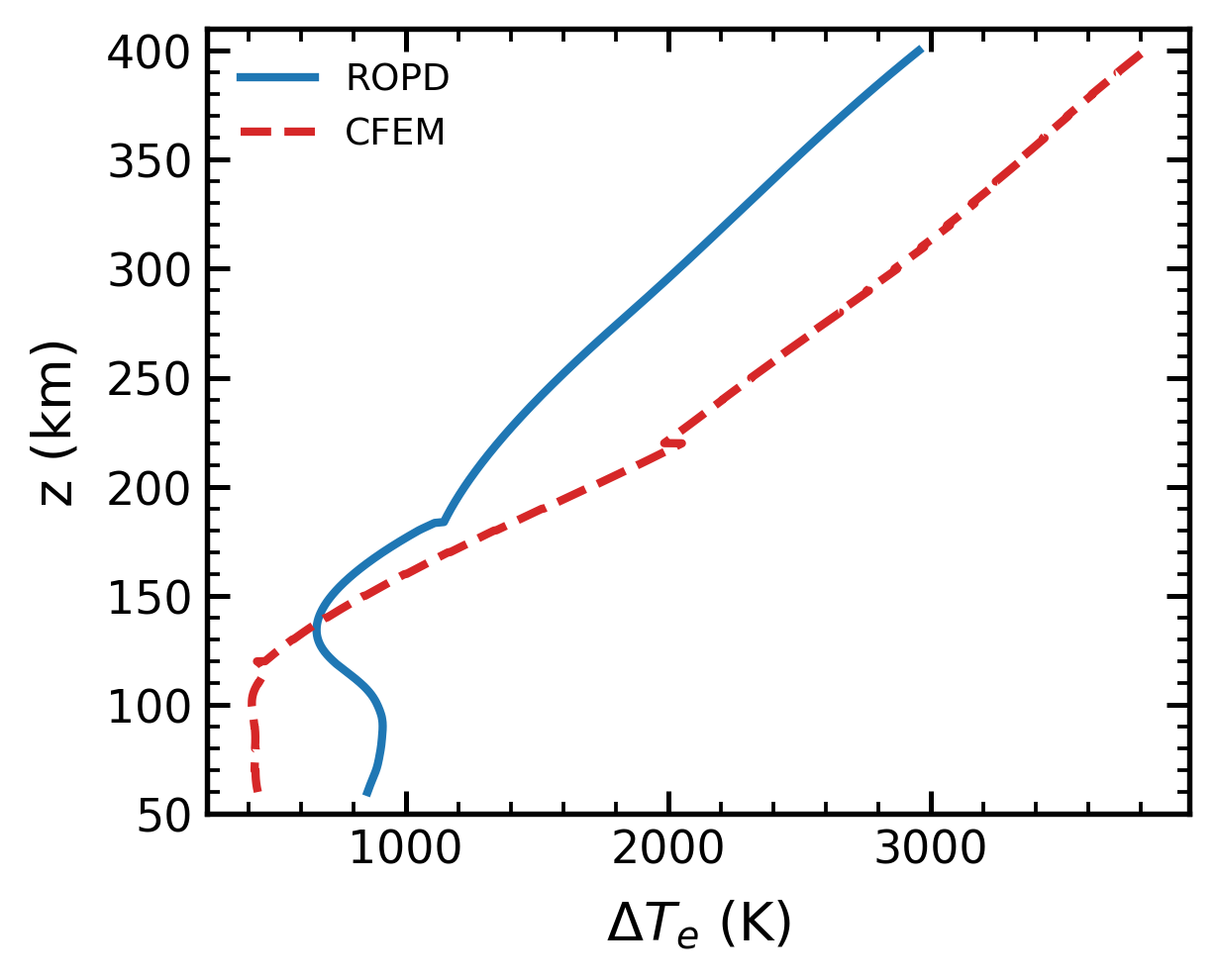}
    \caption{Nonlinear plasma-response estimates in the 2.45 GHz analysis: (a) density perturbation and (b) electron-temperature perturbation from the ROPD reference and CFEM calculation.}
    \label{fig:nonlinear_matlab_comsol}
\end{figure}

Figure~\ref{fig:nonlinear_matlab_comsol} compares the ROPD estimate with the corresponding CFEM result for the nonlinear plasma response. The comparison is intended as a trend-level diagnostic rather than a pointwise validation, because the ROPD response is driven by a one-dimensional intensity estimate based on Gaussian-beam spreading and WKB attenuation, whereas the CFEM response is driven by the local two-dimensional full-wave field coupled to the temperature-dependent dielectric model. The CFEM electron-temperature curve over the 200--250 km transition interval is presented after a local smoothing treatment that reduces numerical artifacts without changing the overall altitude trend.

The two routes are expected to identify similar response channels and altitude localization, but not to coincide point by point. The ROPD calculation uses the reconstructed background profiles, a Gaussian-beam spreading model with WKB absorption, and local analytic expressions for the ponderomotive density response and electron heating. In contrast, the CFEM solution uses the local full-wave electric field from the axisymmetric model, updates the dielectric properties through the neural-network thermal and collision-frequency closure, and includes radial field redistribution, diffraction, near-axis field behavior, and inter-segment field transfer. These effects change the local value of $|\mathbf{E}|^2$ that drives both $Q_h$ and the ponderomotive term, so the nonlinear response evaluated from CFEM need not reproduce the ROPD estimate in amplitude even when both models use the same background ionospheric input. The amplitude difference is therefore a consequence of the model hierarchy rather than an independent indication of a different background ionosphere, and the comparison in Figure~\ref{fig:nonlinear_matlab_comsol} should be read as a cross-model consistency check for the activated response channels and their broad altitude localization.

The most physically significant feature of Figure~\ref{fig:nonlinear_matlab_comsol}(b) is that the electron-temperature perturbation increases with altitude even though the volumetric heating rate decreases: energy is deposited predominantly near 95 km, yet the temperature rises most strongly in the F region, reaching the path maximum of 3815 K near the 400 km entrance. This apparent paradox is resolved by the steady-state balance of Equation~(14). Every cooling channel included in the balance---elastic electron-neutral collisions, rotational and vibrational excitation of $\mathrm{N}_2$ and $\mathrm{O}_2$, and fine-structure excitation of atomic O---scales with the neutral density, which falls by many orders of magnitude between 95 km and 400 km. The F-region electron gas is therefore nearly calorimetrically isolated in the local energy-balance description adopted here: even a heating rate that is orders of magnitude below the E-region value drives a kilokelvin-scale temperature rise because the cooling bottleneck, not the heating strength, sets the response. The practical consequence is a spatial decoupling that a single attenuation number cannot capture: the altitude band that absorbs the beam energy (collision-controlled, near 95 km) is not the altitude band that exhibits the strongest plasma modification (cooling-controlled, in the F region).

The ponderomotive density perturbation in Figure~\ref{fig:nonlinear_matlab_comsol}(a) remains very small along the entire path. This is expected from the scaling of Equations~(11) and (12): at the field amplitudes reached by a 1 GW beam of kilometer-scale radius, the ponderomotive potential $\Phi_p$ remains many orders of magnitude below the electron thermal energy $k_B T_e$, so the Boltzmann response stays in its linear regime, $|\Delta N_e/N_e|\approx\Phi_p/(k_B T_e)\ll1$. The beam therefore operates far below the cavitation and self-focusing thresholds identified in earlier modification studies \cite{Thome1974Striations,Shinohara1995SelfFocusing}, and the density channel acts only as a weak refractive perturbation rather than as a feedback strong enough to restructure the beam.

\subsubsection{Beam Radial Profile and Cascaded Propagation}
\begin{figure}[H]
    \centering
    \resetpanels
    \subpanel{0.48\textwidth}{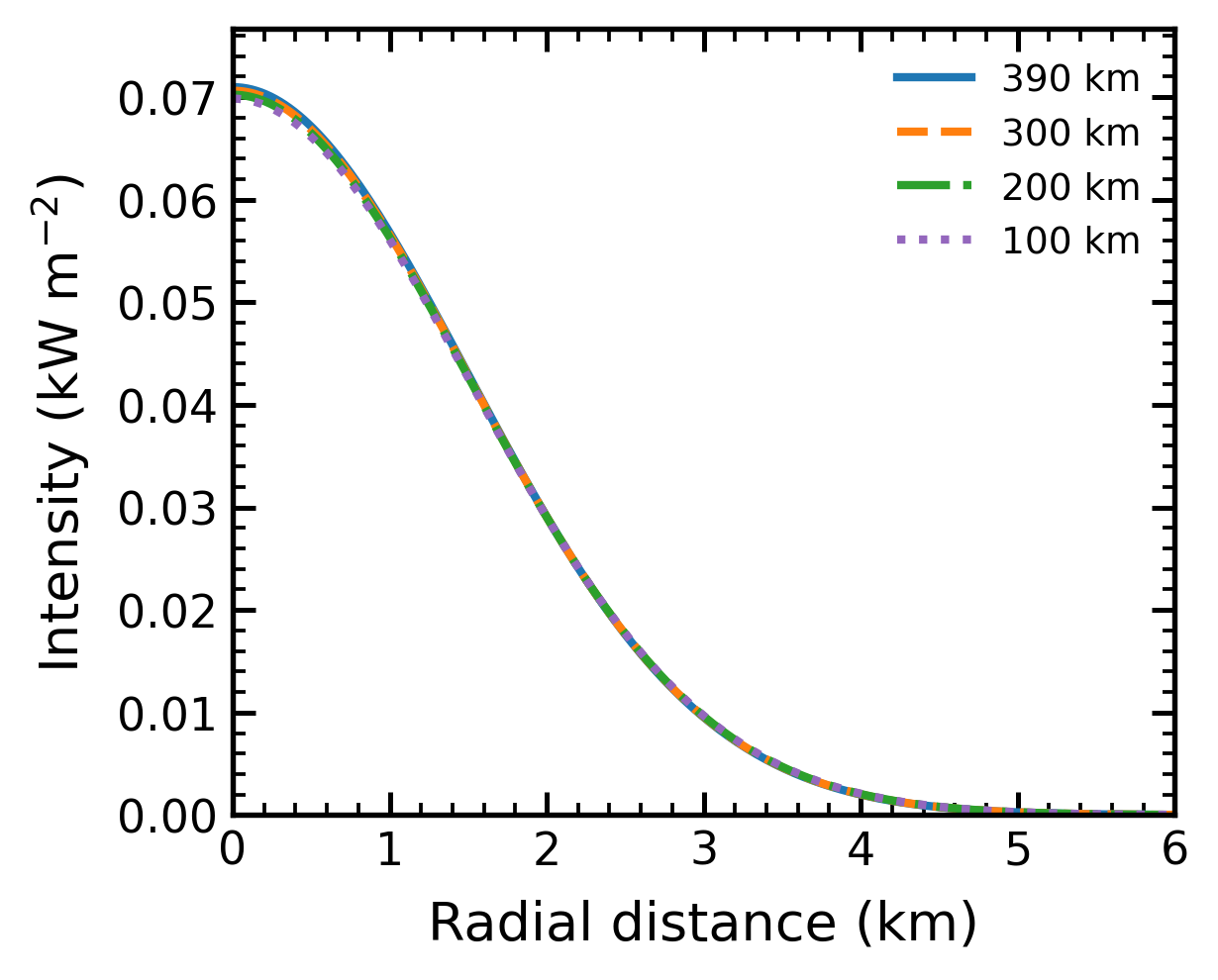}
    \hfill
    \subpanel{0.48\textwidth}{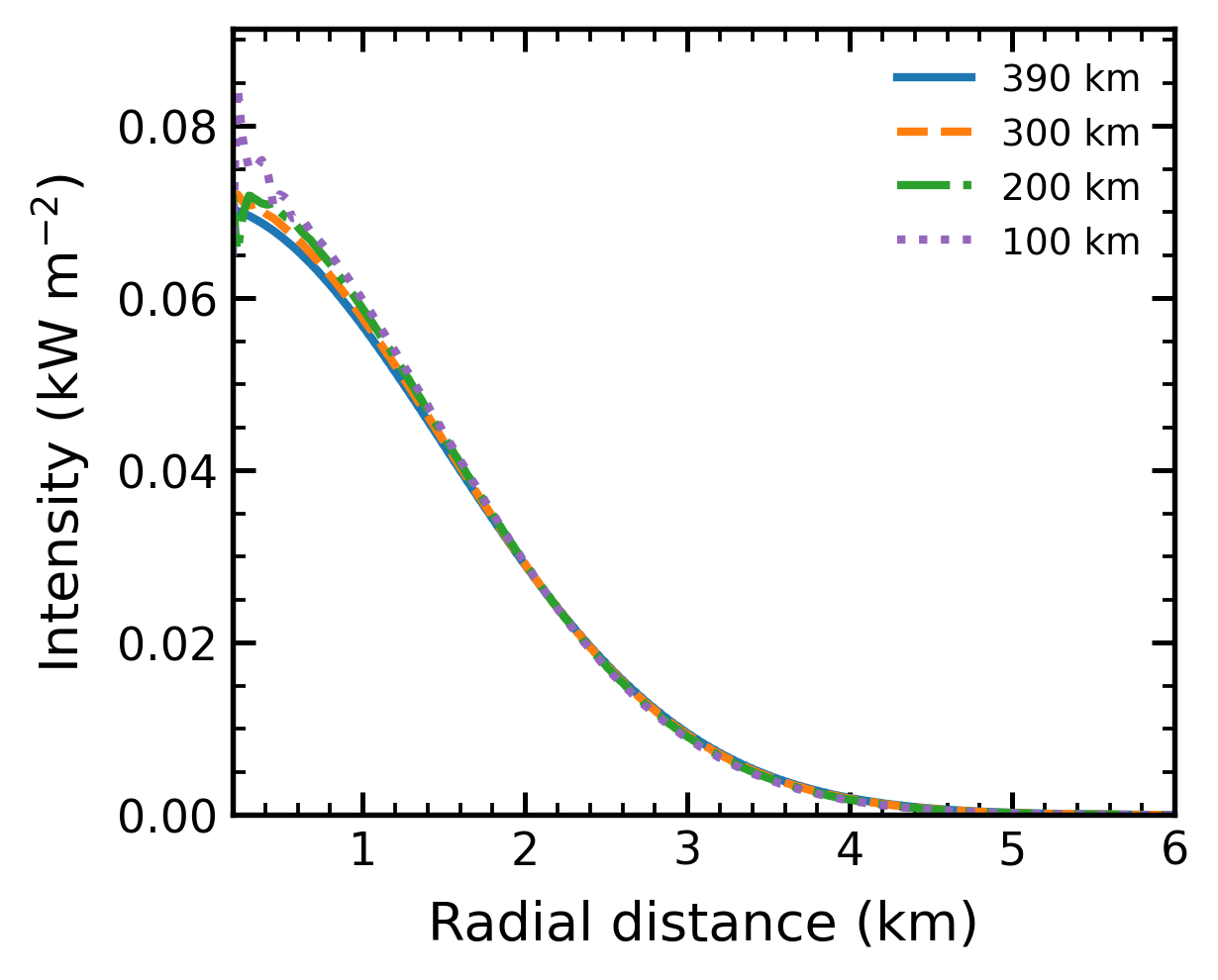}
    \caption{Radial redistribution of beam intensity in the 2.45 GHz analysis: (a) ROPD Gaussian-beam reference and (b) CFEM cascade.}
    \label{fig:beam_radial_comparison}
\end{figure}

\begin{figure}[H]
    \centering
    \resetpanels
    \subpanel{0.32\textwidth}{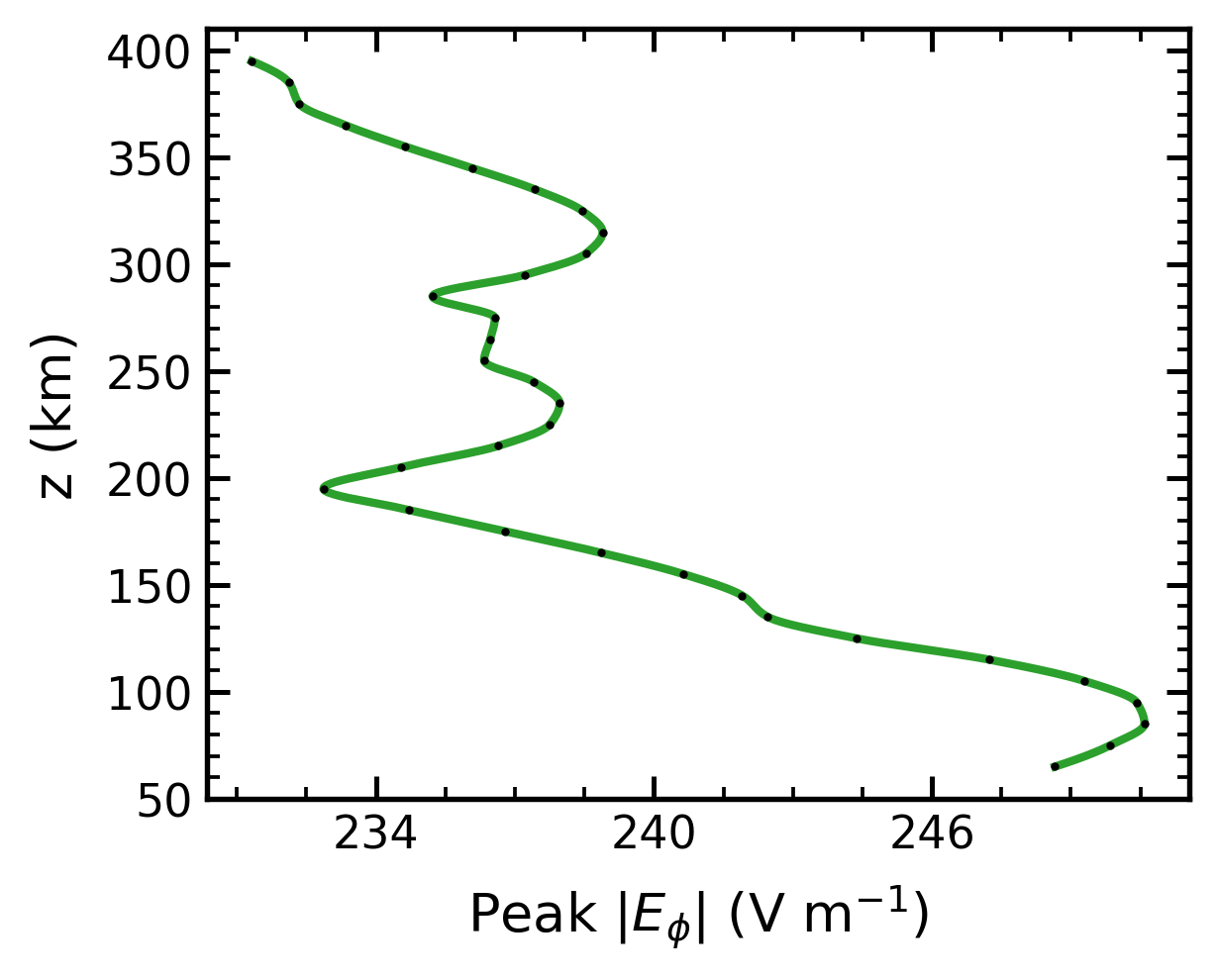}
    \hfill
    \subpanel{0.32\textwidth}{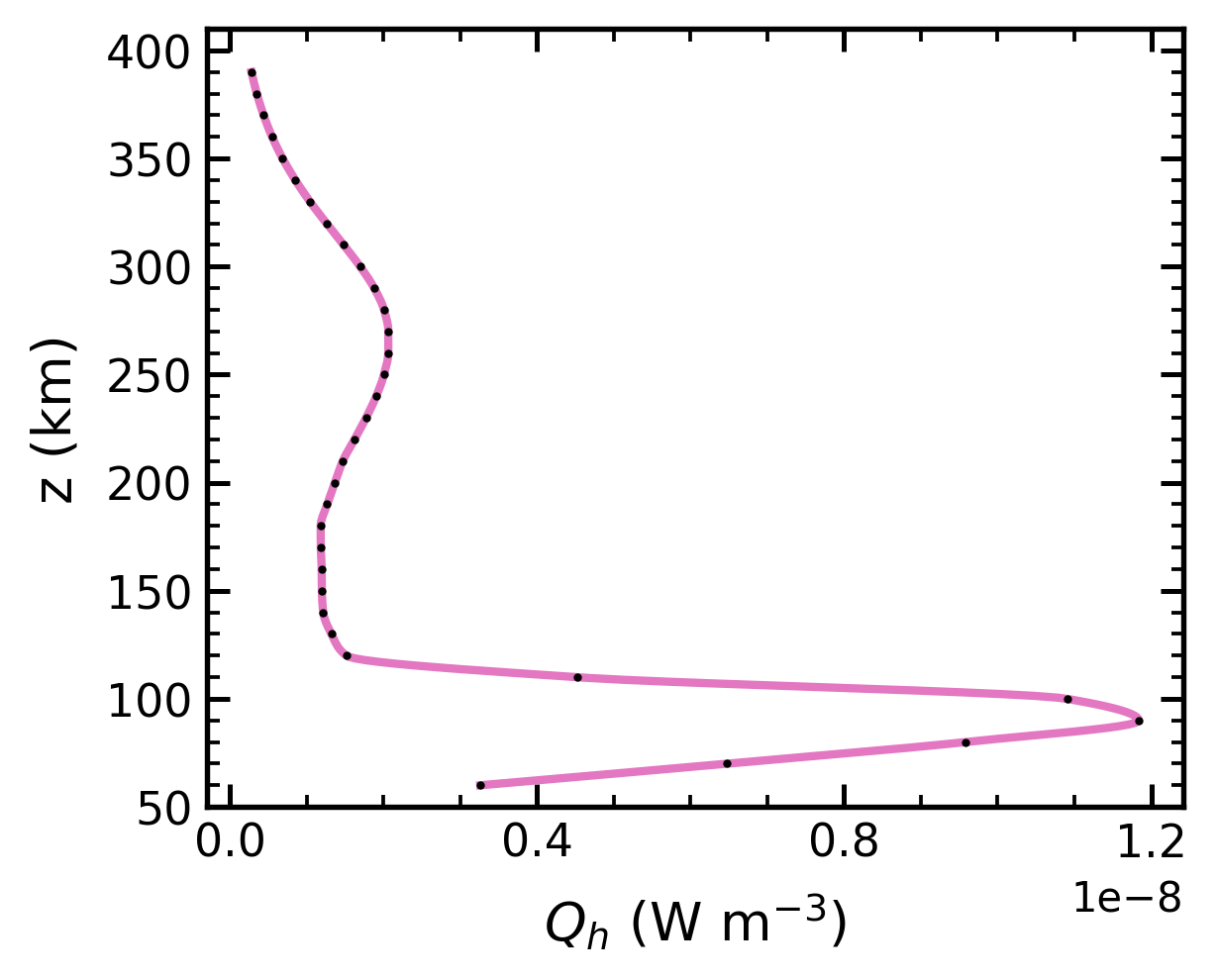}
    \hfill
    \subpanel{0.32\textwidth}{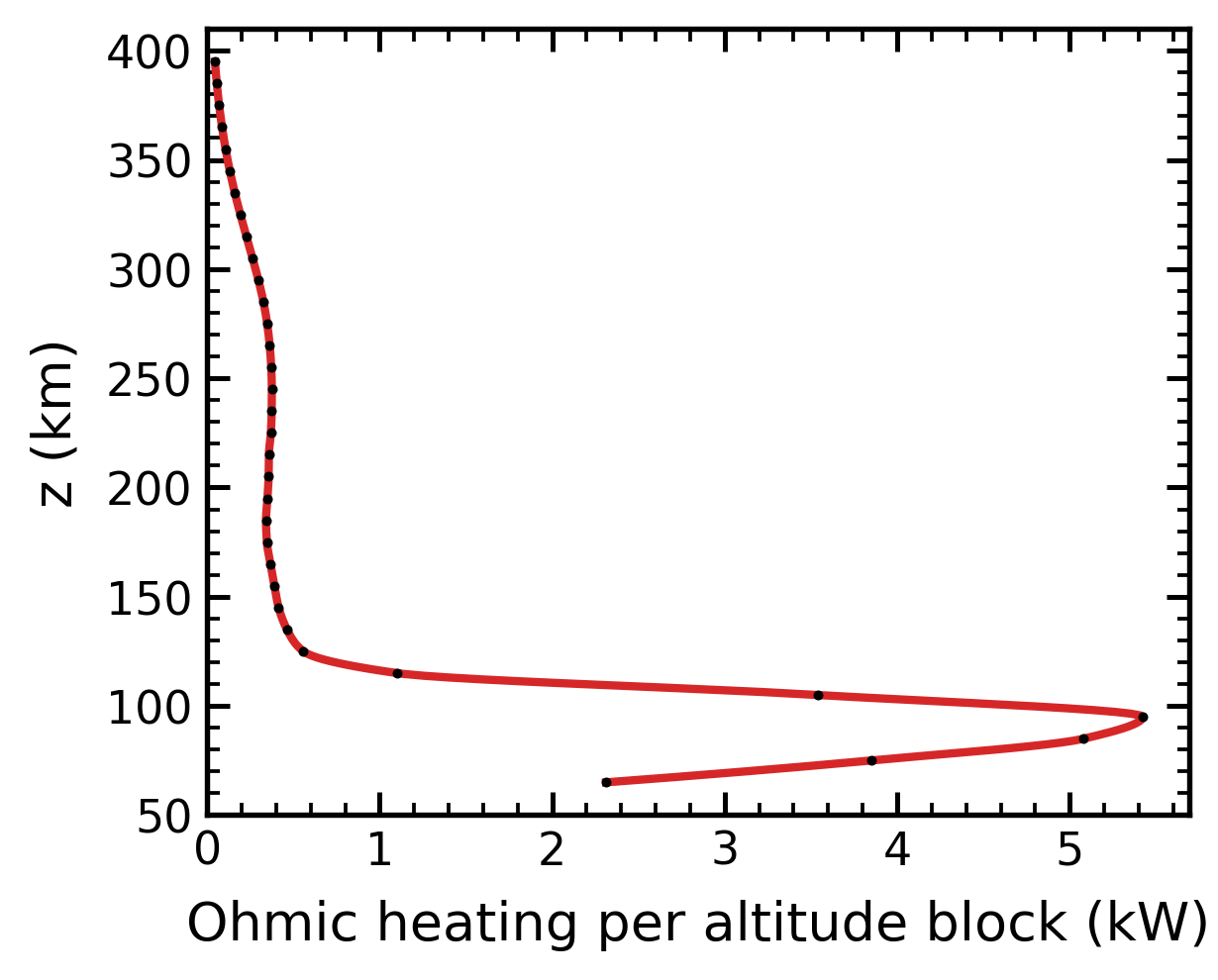}
    \caption{Full-wave cascade evolution in the 2.45 GHz analysis: (a) peak electric-field magnitude, (b) centerline heating rate, and (c) altitude-block integrated Ohmic heating.}
    \label{fig:comsol_cascade}
\end{figure}

Figure~\ref{fig:beam_radial_comparison} highlights the radial beam structure predicted by the Gaussian reference and by the CFEM cascade. In the CFEM panel, the near-axis interval $0\leq r<0.2$ km is omitted because the axisymmetric azimuthal-field regularity condition affects the displayed samples there; this plotting choice does not modify the cascade power-transfer calculation. Figure~\ref{fig:comsol_cascade} then summarizes the segmented propagation evolution and the altitude-block distribution of the volume-integrated Ohmic heating.

In Figure~\ref{fig:comsol_cascade}, each propagation block is evaluated after normalizing the reconstructed incident field to 1 GW. The plotted peak electric-field and centerline heating profiles provide the local field quantities that drive the nonlinear plasma response, while the altitude-block heating panel is obtained by volume integration of $Q_h$ over each 10 km portion of the path. The resulting 2.45 GHz thermal deposition is concentrated in the lower part of the path, with the largest block contribution occurring near 95 km altitude, in agreement with the $N_e\nu$ scaling argument of Section~3.1.1: the altitude-block heating curve in Figure~\ref{fig:comsol_cascade}(c) is essentially a map of where electrons and collisions coexist. Above roughly 120 km the per-block deposition falls to a low, slowly varying level, while the blocks between 80 and 120 km dominate the 29.4 kW total. The peak-field evolution in Figure~\ref{fig:comsol_cascade}(a) is governed primarily by diffraction spreading of the finite beam, confirming that at these field levels nonlinear refractive feedback perturbs, but does not restructure, the beam envelope.

\subsection{For 5.8 GHz Propagation}
The 5.8 GHz analysis follows the same organization as the 2.45 GHz calculation: background and loss-related profiles, nonlinear plasma response, radial beam structure, and segmented propagation evolution.

\subsubsection{Background Profiles and Cross-Model Consistency}
\begin{figure}[H]
    \centering
    \resetpanels
    \subpanel{0.32\textwidth}{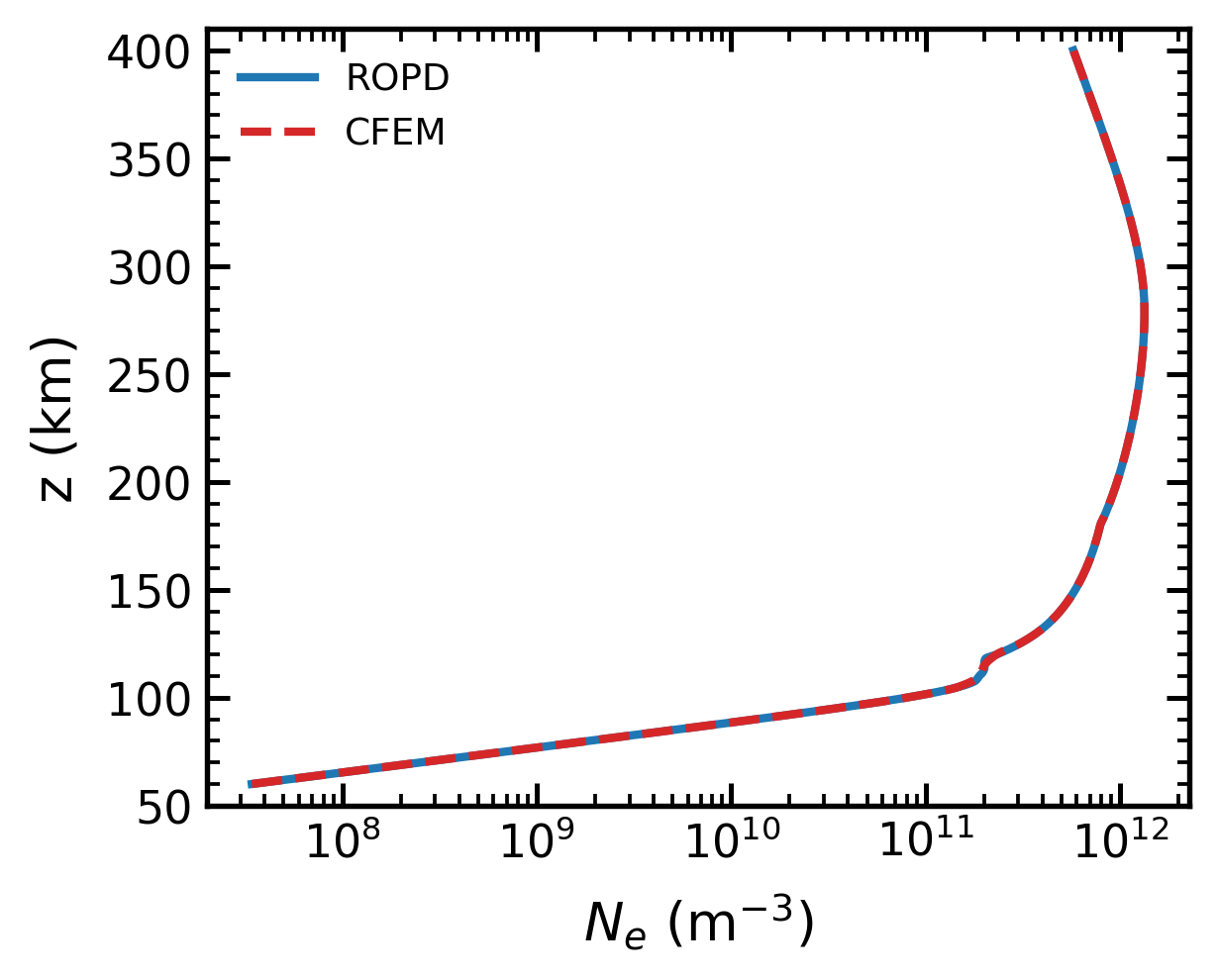}
    \hfill
    \subpanel{0.32\textwidth}{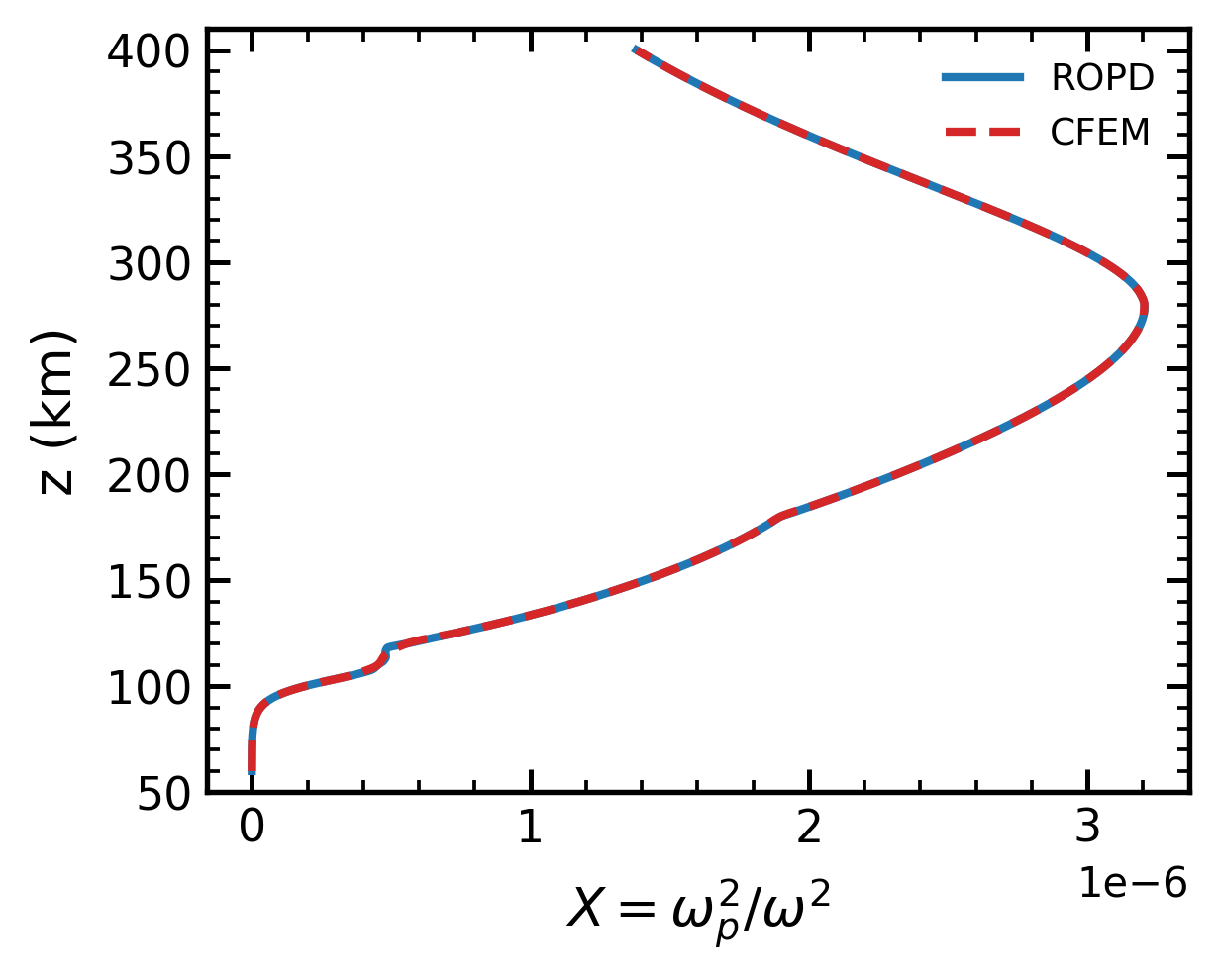}
    \hfill
    \subpanel{0.32\textwidth}{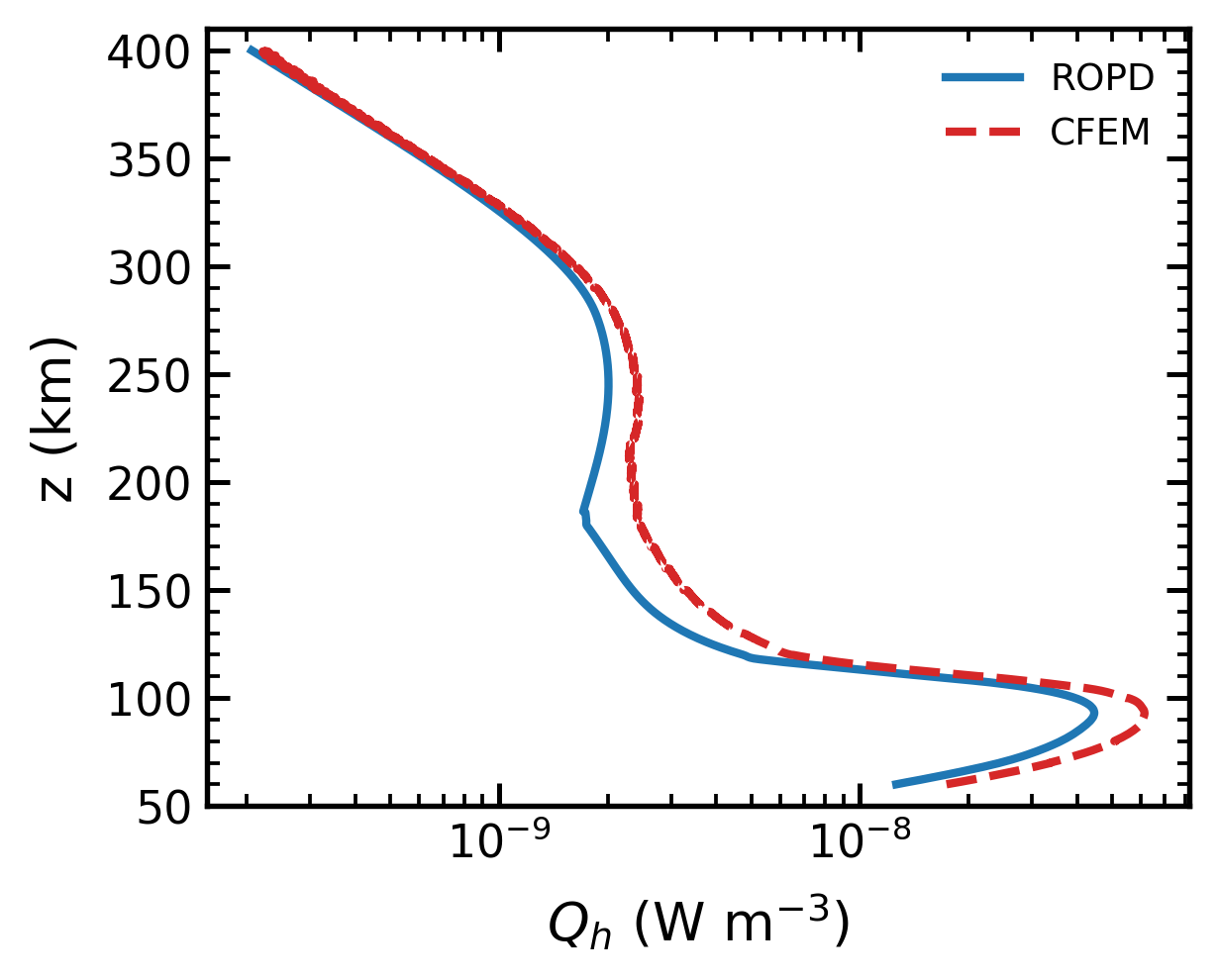}
    \caption{Background and loss-related profiles for the 5.8 GHz analysis: (a) electron density, (b) propagation parameter, and (c) Ohmic heating rate.}
    \label{fig:profile_matlab_comsol_5p8}
\end{figure}

Figure~\ref{fig:profile_matlab_comsol_5p8} summarizes the background and loss-related profiles used in the 5.8 GHz analysis. The electron-density profile is unchanged because it is determined by the same IRI input, while the propagation parameter $X=(\omega_p/\omega)^2$ is smaller than in the 2.45 GHz analysis because of the larger microwave frequency. The Ohmic-heating comparison retains the same ROPD--CFEM interpretation as above: ROPD provides the one-dimensional diagnostic trend, whereas the CFEM result reflects the field distribution obtained after diffraction, boundary absorption, and nonlinear material feedback. The altitude localization of the heating is unchanged, since the $N_e(z)\,\nu(z)$ product that controls the deposition profile is a property of the background medium rather than of the beam.

\subsubsection{Nonlinear Heating and Plasma Response}
\begin{figure}[H]
    \centering
    \resetpanels
    \subpanel{0.48\textwidth}{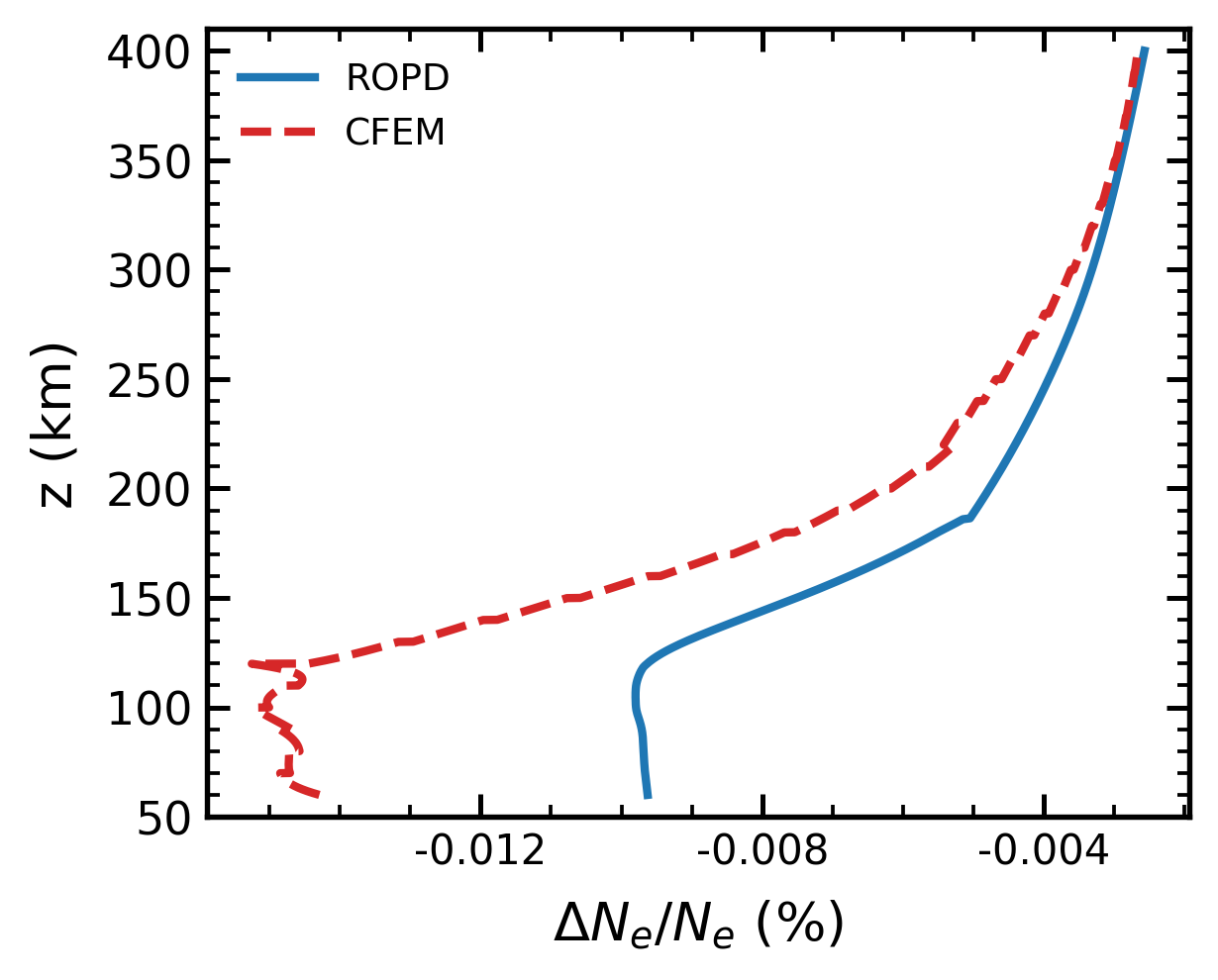}
    \hfill
    \subpanel{0.48\textwidth}{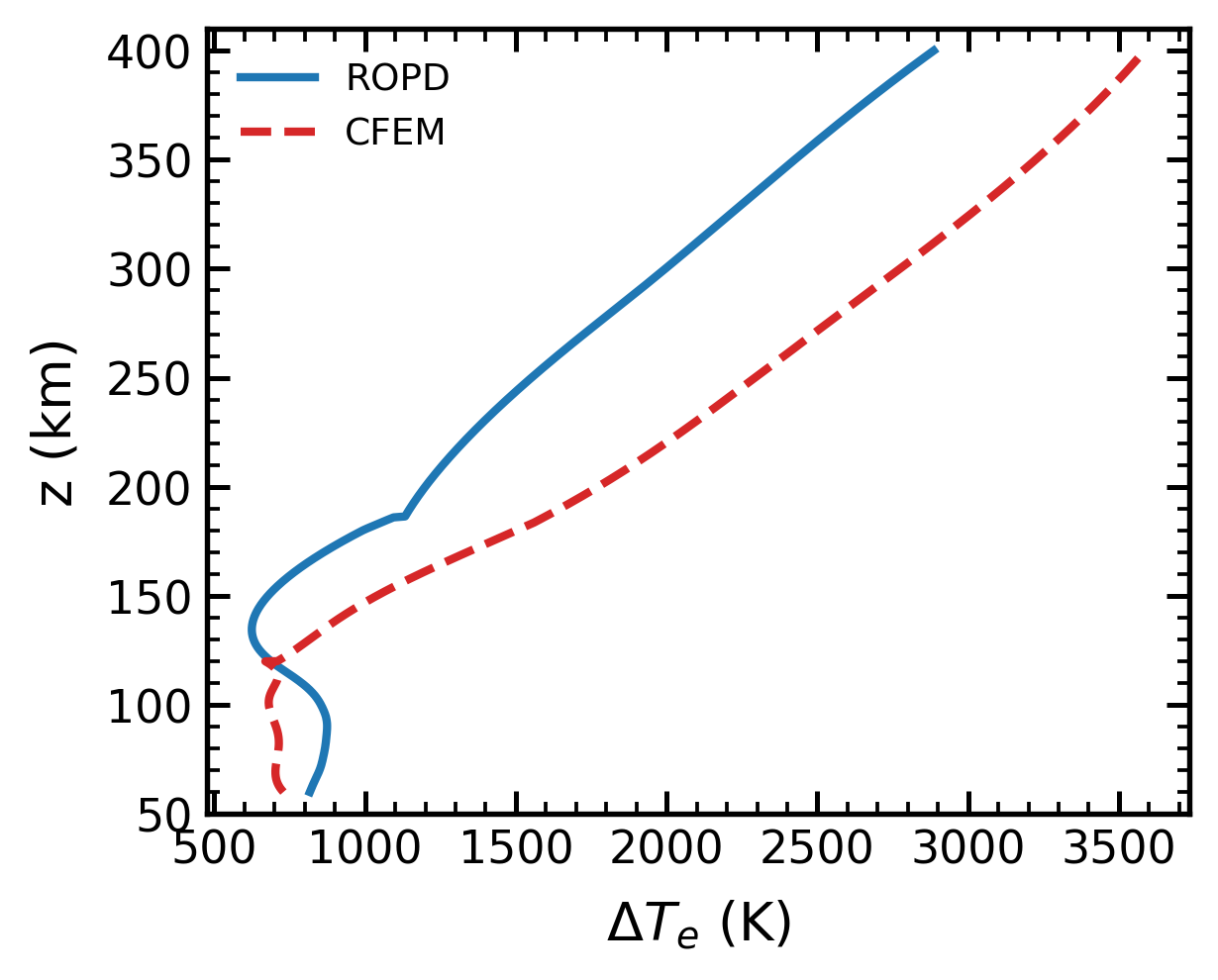}
    \caption{Nonlinear plasma-response estimates in the 5.8 GHz analysis: (a) density perturbation and (b) electron-temperature perturbation from the ROPD reference and CFEM calculation.}
    \label{fig:nonlinear_matlab_comsol_5p8}
\end{figure}

The nonlinear plasma response at 5.8 GHz is shown in Figure~\ref{fig:nonlinear_matlab_comsol_5p8}. As in the 2.45 GHz case, the ROPD curve should be viewed as a trend estimate rather than a pointwise reference for the CFEM result. The density perturbation and electron-temperature perturbation are both driven by the local electric-field magnitude in the CFEM solution, so their amplitudes are affected by frequency-dependent beam evolution, radial field redistribution, and nonlinear dielectric feedback. Compared with the 2.45 GHz case, the larger carrier frequency reduces the plasma propagation parameter and weakens the ponderomotive potential for a fixed field amplitude; at the same time, the narrower 5.8 GHz beam carries a higher on-axis intensity for the same 1 GW input. The net outcome of these competing scalings---a larger peak density-depletion magnitude but a slightly smaller maximum electron-temperature perturbation than at 2.45 GHz---is quantified and interpreted in Section~3.3, and is consistent with the summary indicators in Table~\ref{tab:dual_frequency_summary}. The CFEM $\Delta T_e$ curve around 200--250 km is presented after a local smoothing treatment that reduces numerical artifacts without changing the overall altitude trend.

\subsubsection{Beam Radial Profile and Cascaded Propagation}
\begin{figure}[H]
    \centering
    \resetpanels
    \subpanel{0.48\textwidth}{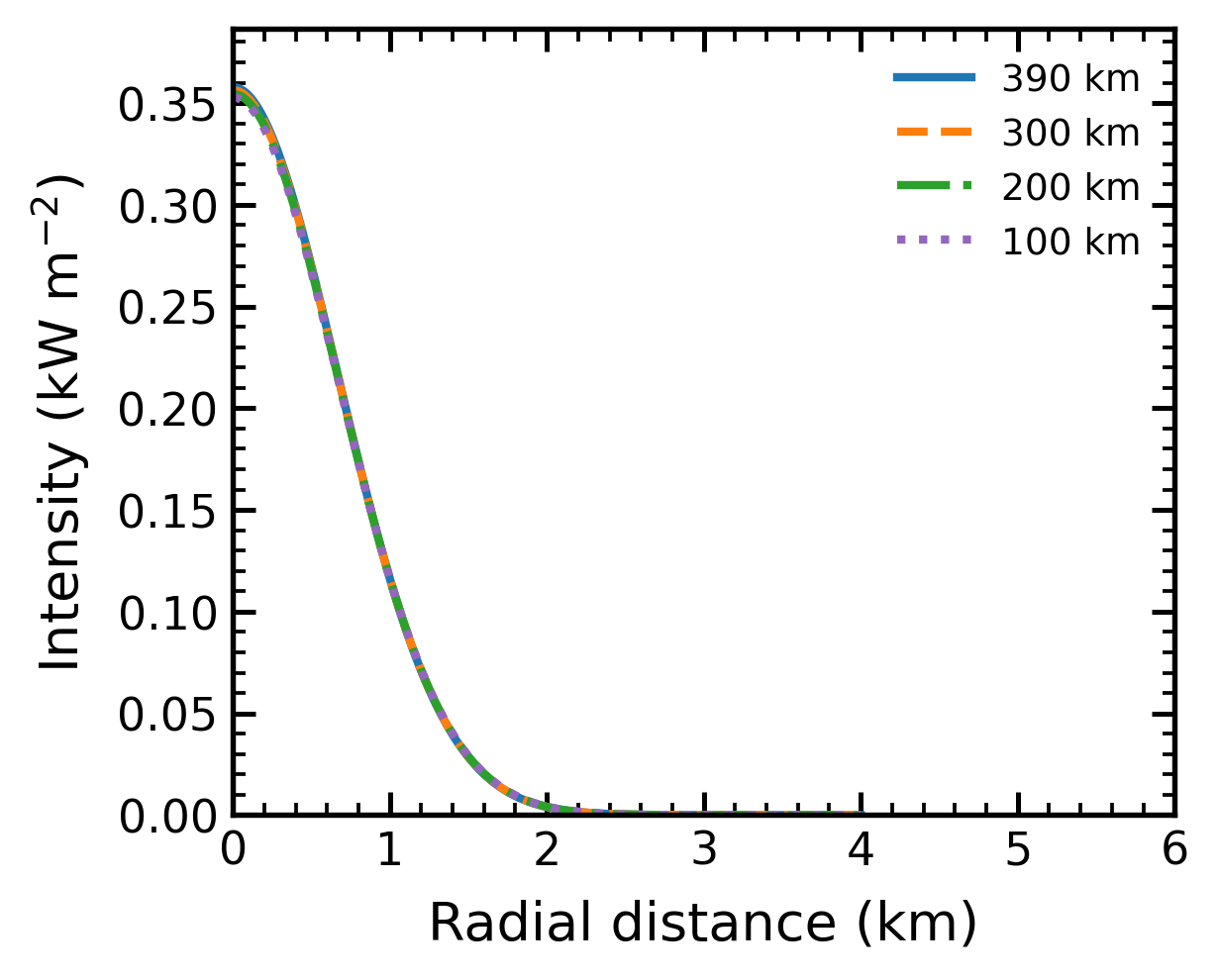}
    \hfill
    \subpanel{0.48\textwidth}{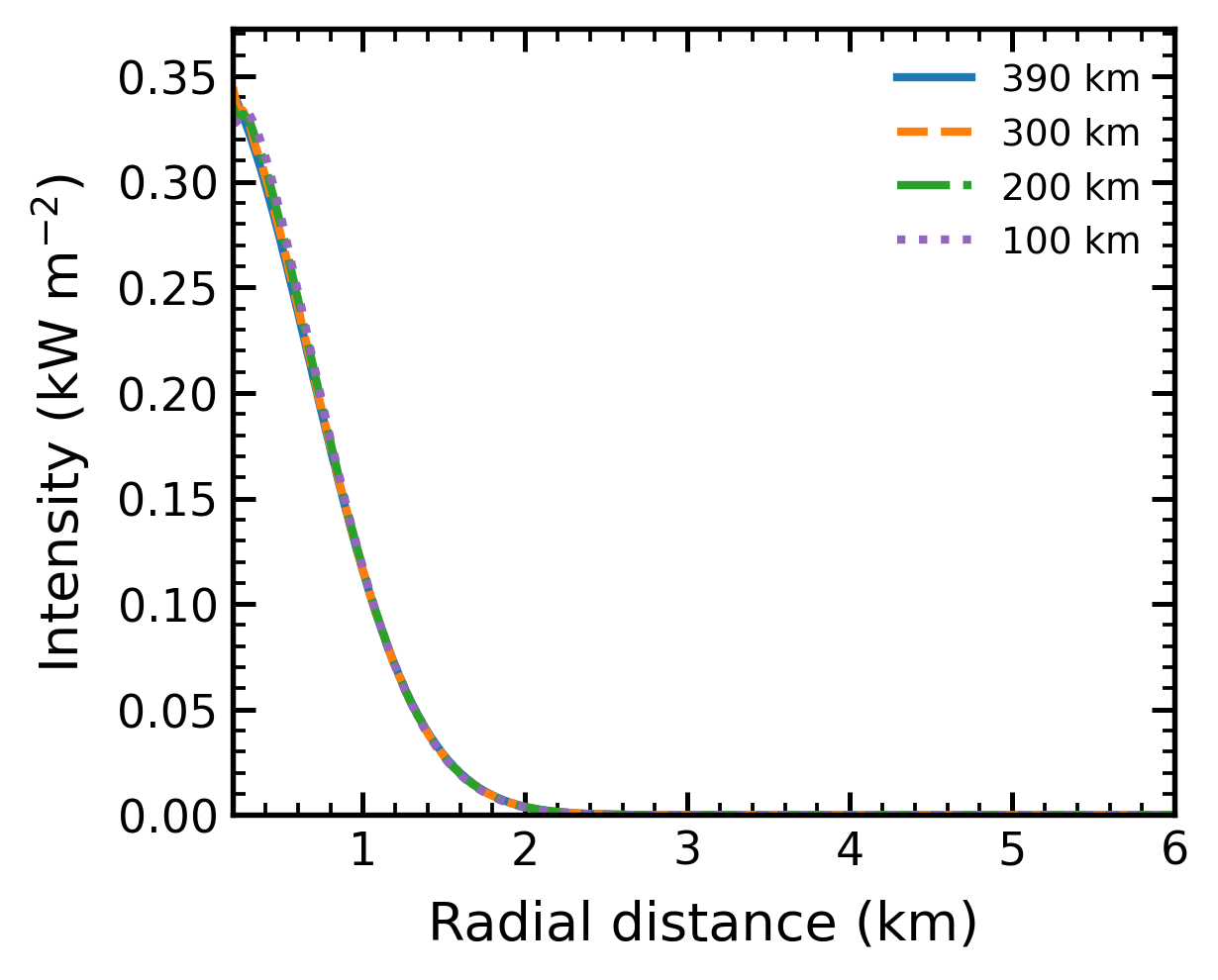}
    \caption{Radial redistribution of beam intensity in the 5.8 GHz analysis: (a) ROPD Gaussian-beam reference and (b) CFEM cascade.}
    \label{fig:beam_radial_comparison_5p8}
\end{figure}

Figure~\ref{fig:beam_radial_comparison_5p8} highlights the radial beam structure predicted by the Gaussian reference and by the CFEM cascade. Compared with the 2.45 GHz radial profiles in Figure~\ref{fig:beam_radial_comparison}, the 5.8 GHz profiles have a higher on-axis intensity because the shorter wavelength produces a narrower free-space Gaussian beam at the entrance altitude under the same transmitted power and aperture setting. The panels are not expected to match point by point because the ROPD result is a reference estimate, while the CFEM solution includes radial redistribution and inter-segment complex-field transfer. As in the 2.45 GHz CFEM radial panel, the near-axis interval $0\leq r<0.2$ km is omitted only in the plotted CFEM profile to avoid regularity-condition effects, without changing the integrated cascade-power calculation.

\begin{figure}[H]
    \centering
    \resetpanels
    \subpanel{0.32\textwidth}{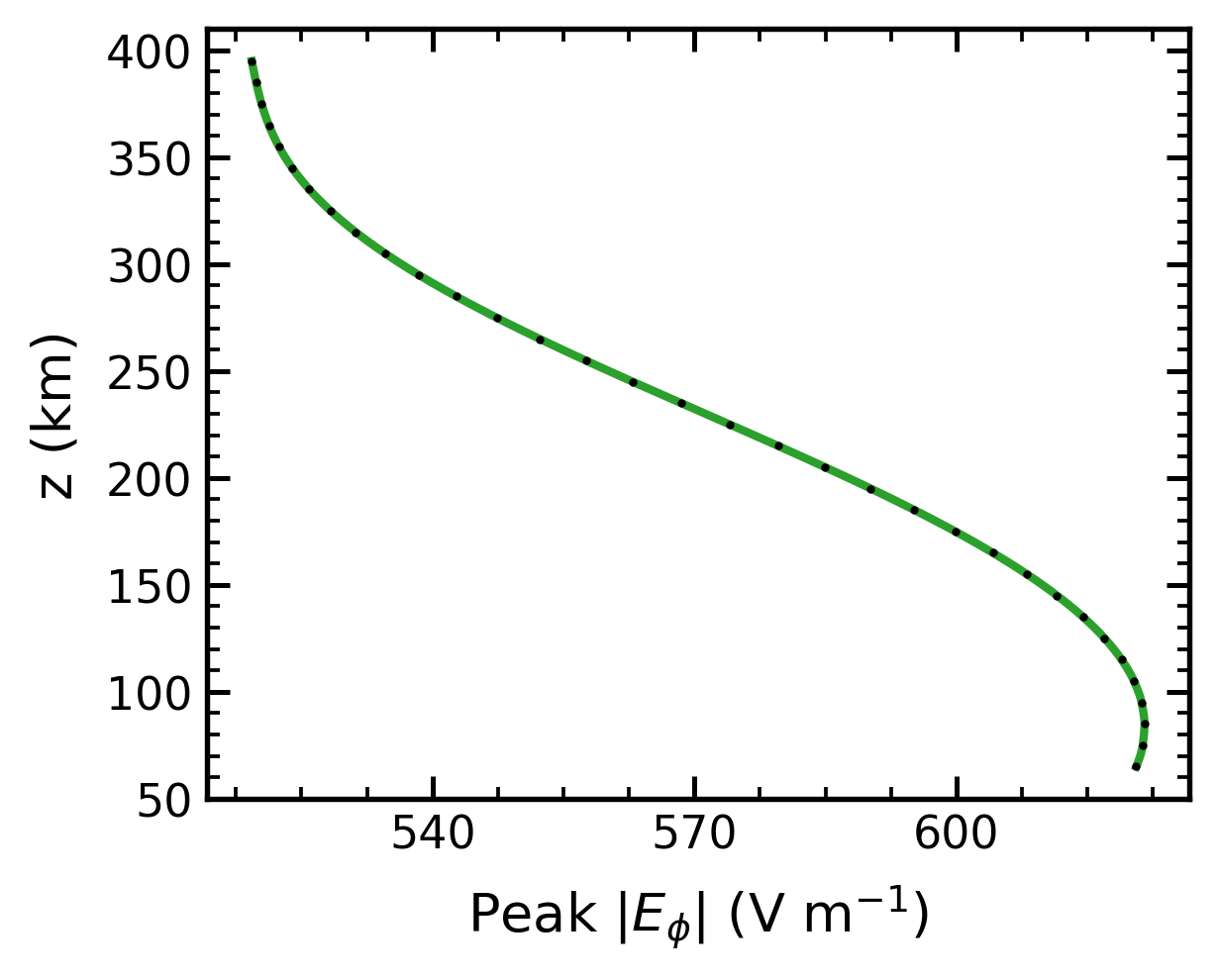}
    \hfill
    \subpanel{0.32\textwidth}{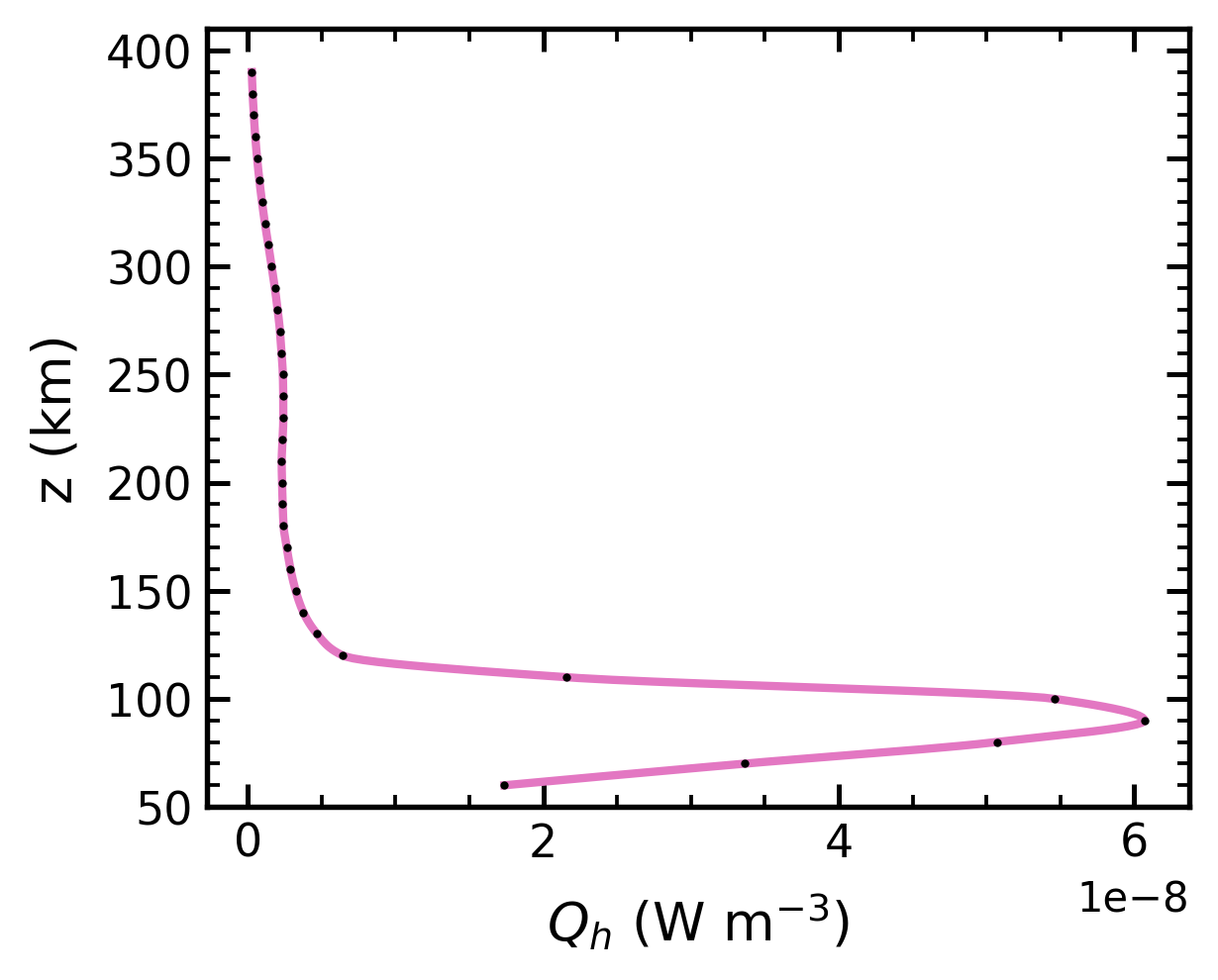}
    \hfill
    \subpanel{0.32\textwidth}{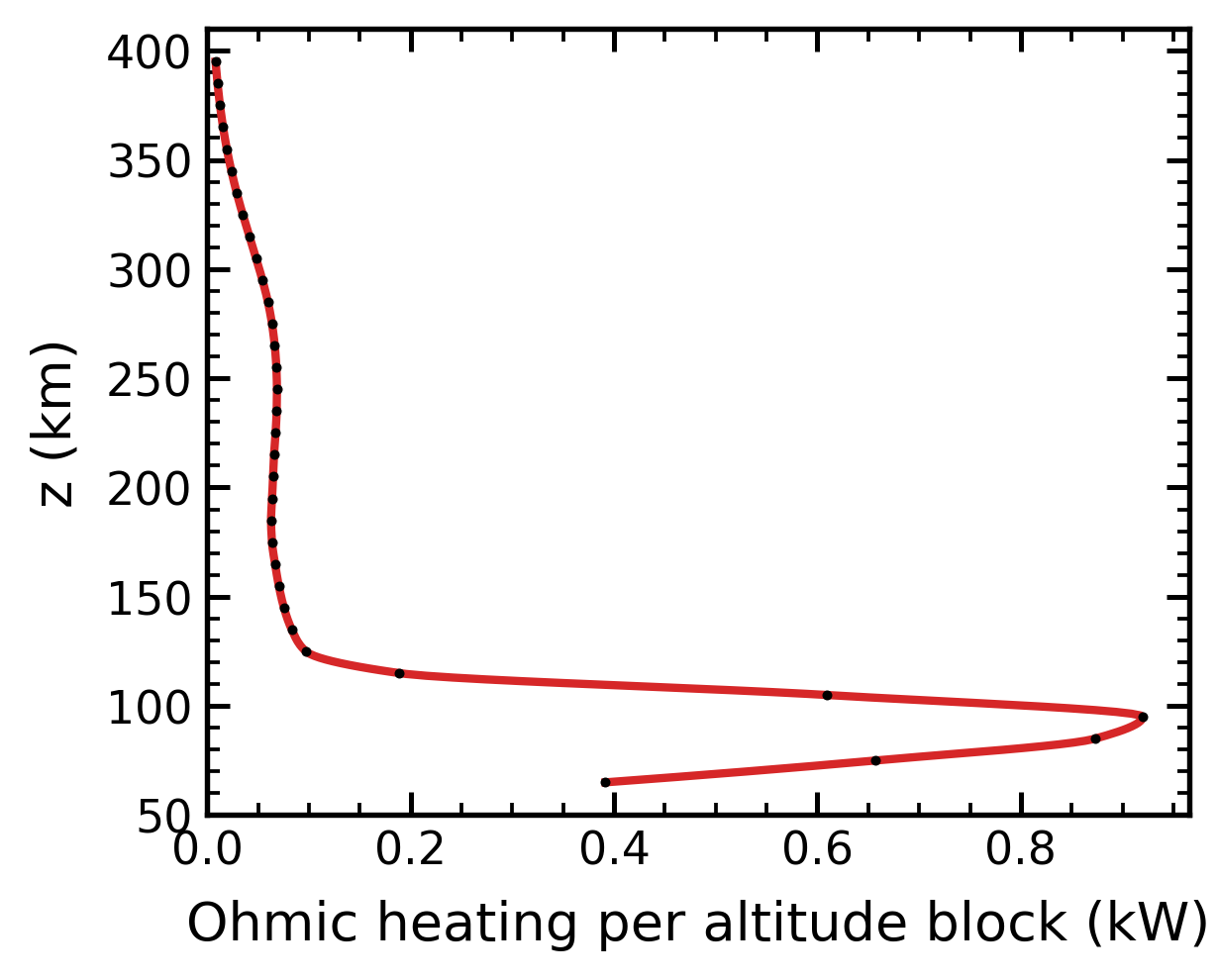}
    \caption{Full-wave cascade evolution in the 5.8 GHz analysis: (a) peak electric-field magnitude, (b) centerline heating rate, and (c) altitude-block integrated Ohmic heating.}
    \label{fig:comsol_cascade_5p8}
\end{figure}

The segmented propagation evolution is summarized in Figure~\ref{fig:comsol_cascade_5p8}. As in the 2.45 GHz analysis, each local segment is evaluated with a 1 GW normalized incident field reconstructed from the preceding complex field. The plotted peak electric-field, centerline heating, and altitude-block heating profiles provide complementary views of the CFEM quantities that drive the nonlinear plasma response. The 5.8 GHz altitude-block heating remains smaller than the 2.45 GHz value but follows the same lower-altitude concentration, with the largest block contribution near 95 km altitude.

\subsection{Cross-Band Comparison and Frequency Scaling}
{\sloppy The main CFEM indicators obtained from the evaluated fields are summarized in Table~\ref{tab:dual_frequency_summary}. The normalized-loss entry reports the dimensionless cascade transfer loss used for comparing the two frequency cases. The total Ohmic heating entry is computed independently from the volumetric $Q_h$ deposition over the 34 altitude blocks and is consistent with the altitude-block heating panels in Figures~\ref{fig:comsol_cascade} and \ref{fig:comsol_cascade_5p8}. The heating, temperature-perturbation, and density-depletion entries report extrema over the modeled altitude interval. These quantities are extracted from the underlying CFEM fields; local presentation treatments and near-axis omissions in selected radial-profile plots affect only visualization and not the volume-integrated heating calculation.\par}
\begin{table}[H]
\centering
\caption{CFEM summary indicators for the two normalized transmission cases.}
\label{tab:dual_frequency_summary}
\scriptsize
\begin{tabular}{lccccc}
\hline
$f$ (GHz) & Normalized loss & Total heating (kW) & Max. $Q_h$ (W m$^{-3}$) & Max. $\Delta T_e$ (K) & Max. $|\Delta N_e/N_e|$ (\%) \\
\hline
2.45 & $9.0\times10^{-5}$ & 29.416 & $1.18\times10^{-8}$ & 3815 & 0.00565 \\
5.8 & $3.3\times10^{-4}$ & 5.113 & $6.07\times10^{-8}$ & 3585 & 0.01525 \\
\hline
\end{tabular}
\end{table}

Three quantitative relations in Table~\ref{tab:dual_frequency_summary} carry direct physical meaning.

First, the total heat deposition follows the collisional-absorption frequency scaling almost exactly. In the regime $\nu\ll\omega$, Equation~(13) predicts $Q_h\propto\omega^{-2}$ for a given field intensity, and because the cross-sectional integral of $|\mathbf{E}|^2$ is fixed by the common 1 GW input power, the volume-integrated deposition should scale as $\omega^{-2}$ if the background plasma were unperturbed. The computed ratio, $29.4/5.11\approx5.75$, lies within a few percent of the ideal value $(5.8/2.45)^2\approx5.60$. This close agreement demonstrates that the bulk absorption remains in the linear collisional regime even for a GW-class beam, with the small residual excess attributable to the nonlinear thermal feedback---the heated electrons increase $\nu_{en}$ through the temperature-dependent collision formulas---and to the slightly different field distributions of the two cascades. For engineering purposes, this validates a simple design rule: for a fixed background ionosphere, the fractional ionospheric Ohmic loss of an SSPS link scales as $f^{-2}$, so moving from 2.45 GHz to 5.8 GHz buys a further factor of about 5.6 reduction in an already negligible loss.

Second, the maximum electron-temperature perturbations are nearly equal (3815 K versus 3585 K) even though the ohmic-heating efficiency per unit intensity differs by a factor of 5.6 between the bands. Two compensating effects produce this near-parity. The shorter 5.8 GHz wavelength yields a narrower vacuum beam and therefore a higher on-axis $|\mathbf{E}|^2$ for the same transmitted power, which largely offsets the $\omega^{-2}$ penalty in the local heating rate that drives the F-region temperature response. In addition, the steady-state temperature is a strongly sublinear function of the heating rate, because the cooling terms in Equation~(14) rise steeply with $T_e$; large differences in $Q_h$ therefore map onto much smaller differences in the equilibrium temperature. The F-region thermal perturbation is thus a robust feature of GW-class transmission in this band pair, rather than a quantity that can be engineered away by the choice between the two candidate frequencies.

Third, the maximum density depletion is larger at 5.8 GHz (0.01525\%) than at 2.45 GHz (0.00565\%), which at first sight contradicts the $\Phi_p\propto\omega^{-2}$ scaling of the ponderomotive potential in Equation~(11). The resolution is that the local scaling argument is incomplete at the propagation scale. For beams launched from the same aperture, the far-field on-axis intensity scales approximately as $\omega^2$, so the leading-order ponderomotive response is nominally frequency-independent; the factor of about 2.7 difference obtained from the full calculation therefore reflects genuinely propagation-scale physics---the narrower, less-diffracting 5.8 GHz beam maintains its high on-axis intensity over a longer portion of the path, and its slightly lower electron temperature deepens the Boltzmann response of Equation~(12). This inversion of the naive single-point scaling is a concrete demonstration of why full-path modeling is required: band comparisons based on local nonlinear scalings alone can rank the candidate frequencies incorrectly. In absolute terms, however, both depletions remain below 0.02\%, so the density channel stays perturbative at both frequencies.

\section{Conclusion}
\label{sec:conclusion}
This paper presented a coupled CFEM--ROPD framework for evaluating nonlinear SSPS microwave propagation through the ionosphere at the two candidate transmission frequencies, 2.45 GHz and 5.8 GHz. The CFEM resolves the 2D-axisymmetric field evolution through a spatial cascade with complex-field transfer, the ROPD reconstructs the same ionospheric inputs and diagnostics for independent consistency checks, and an SNN surrogate replaces the implicit electron thermal-balance calculation with a smooth explicit closure, improving the numerical stability of the nonlinear dielectric update. To the best of our knowledge, this constitutes the first quantitative evaluation of the mutual effect between microwave power transmission and the ionospheric plasma environment obtained with full-path nonlinear modeling: propagation-scale Maxwell simulation is linked to local plasma heating, collision-frequency variation, and density perturbation over the entire 400--60 km path, without repeated local nonlinear root solving inside the field iterations.

Three physical conclusions emerge from the computed results. First, the ionosphere is effectively transparent to the SSPS power budget. The volume-integrated Ohmic deposition---29.4 kW at 2.45 GHz and 5.11 kW at 5.8 GHz for a 1 GW beam---corresponds to fractional collisional losses of order $3\times10^{-5}$ and $5\times10^{-6}$, respectively, and the near-exact $\omega^{-2}$ ratio between the bands shows that the absorption remains in the linear collisional regime with only a small nonlinear thermal correction. Second, the energy deposition and the plasma response are spatially decoupled. The deposition is collision-controlled and concentrates in the lower E region near 95 km, where the product $N_e\nu$ peaks, whereas the electron-temperature perturbation is cooling-controlled and maximizes in the F region, where the collapse of the neutral density removes the cooling channels; kilokelvin-scale electron heating (up to 3815 K) therefore appears precisely where the volumetric heating rate is weakest. Third, the nonlinear channels are active but perturbative. The ponderomotive density depletion remains below 0.02\% at both frequencies---far below self-focusing and cavitation thresholds, consistent with the threshold analyses of Thome, Perkins, and Shinohara \emph{et al.} \cite{Thome1974Striations,Shinohara1995SelfFocusing}---while the localized heating and refractive perturbation nevertheless modify the local field and accumulated phase. These findings agree with the heating picture of Perkins and Roble \cite{Perkins1978Ionospheric} and with MINIX-related evidence that intense microwaves can generate local plasma activity while maintaining high pump-wave transmission \cite{Kaya1986MINIX,Nagatomo1986MINIX,Matsumoto1986Cyclotron,Matsumoto1995Nonlinear,Omura2005SSPS}, but they place those mechanism-level insights, for the first time, in a quantitative full-path propagation context.

For SSPS engineering, the central message is that the ionosphere is a phase problem, not a loss problem. On the link-budget side, ionospheric Ohmic absorption is orders of magnitude below atmospheric, pointing, and conversion losses and can be carried as a fixed $f^{-2}$ correction. On the beam-control side, however, the F-region thermal perturbation and the field-driven refractive-index modification are imprinted on the very volume that a retrodirective pilot signal must traverse: both the uplink pilot and the downlink power beam propagate through a plasma that the power beam itself has modified, so phase-front distortion accumulates along the path and drifts with the slow thermal response of the plasma. Beam-forming and rectenna phase-compensation loops should therefore budget for a slowly varying, altitude-dependent ionospheric phase term rather than assuming a fixed free-space propagation phase. For frequency selection, the results sharpen the trade space: 5.8 GHz reduces both the propagation parameter $X$ and the absorbed power by the $f^{-2}$ factor, but exhibits the larger local density perturbation because its narrower beam sustains higher on-axis intensity along the path---demonstrating that band comparisons must be made with propagation-scale models rather than local scaling arguments. For environmental and regulatory assessment, the most significant signature is not the small absorbed power but the persistent kilokelvin electron-temperature channel through the F region within the beam footprint, which can perturb local plasma parameters along the beam and should be observable by incoherent-scatter radar or GNSS-based total-electron-content monitoring; the deposition near 95 km, though small in absolute terms, is co-located with the D/E-region altitudes that control HF absorption and is therefore the natural focus for lower-ionosphere impact studies.

Several limitations bound the present conclusions and define the path forward. The framework treats a representative vertical-propagation scenario with an isotropic plasma response and an idealized Gaussian-tapered input; it does not resolve kinetic three-wave coupling or electrostatic wave growth of the type emphasized in MINIX-related studies, and the local energy-balance closure omits field-aligned electron heat conduction, which would spread and moderate the F-region temperature response predicted here. Future work should address the geomagnetic-field-dependent anisotropic dielectric response, oblique incidence and realistic phased-array side-lobe structure, time-dependent ionospheric disturbances, and coupled reduced-order or kinetic descriptions where three-wave processes become important. With these extensions, the framework can evolve from a representative propagation analysis into a design and environmental-assessment tool for operational SSPS microwave links.
\section*{Acknowledgments}
\subsection*{General}
The authors gratefully acknowledge Professor Lei Chang for his valuable guidance, support, and helpful suggestions throughout this work. The authors also thank the members of the heliconX research group at Chongqing University for their helpful discussions and assistance during the preparation of this manuscript.

\subsection*{Author Contributions}
Pengan Guo contributed to the conceptualization and methodology of the study, carried out the full numerical modeling, implemented the CFEM and ROPD workflows, analyzed and visualized the results, and wrote the original draft.

Lei Chang proposed the research idea, provided conceptual and methodological guidance, supervised the project, guided the research process, interpreted the results, and reviewed and edited the manuscript.

Yuhan Chen contributed to the literature review and the organization of background materials.

Ya Gao contributed to the literature review and checked the consistency of figures, tables, and related descriptions.

Longshuai Ye assisted with data organization and the preparation of manuscript materials.

Jikai Sun contributed to technical discussions and provided guidance on the numerical-model implementation.

All authors read and approved the final manuscript.

\subsection*{Funding}
{\raggedright This work was supported by the National Natural Science Foundation of China (Grants No. 92271113, 12411540222, 12481540165, 12405274); the Natural Science Foundation Project of Chongqing (Grant No. CSTB2025NSCQ-GPX0725); and the ENN Hydrogen--Boron Fusion Research Fund (Grant No. 2025ENNHB01-011).\par}

\subsection*{Conflicts of Interest}
The authors declare that there is no conflict of interest regarding the publication of this article.

\subsection*{Data Availability}
The data are available from the authors upon reasonable request.

\printbibliography

@book{Schunk2009Ionospheres,
  author    = {Schunk, Robert W. and Nagy, Andrew F.},
  title     = {Ionospheres: Physics, Plasma Physics, and Chemistry},
  edition   = {2nd},
  publisher = {Cambridge University Press},
  year      = {2009},
  address   = {Cambridge, UK},
  doi       = {10.1017/CBO9780511635342},
  url       = {https://doi.org/10.1017/CBO9780511635342}
}

@article{Perkins1978Ionospheric,
  author  = {Perkins, F. W. and Roble, R. G.},
  title   = {Ionospheric heating by radio waves: Predictions for Arecibo and the satellite power station},
  journal = {Journal of Geophysical Research},
  volume  = {83},
  number  = {A4},
  pages   = {1611--1624},
  year    = {1978},
  doi     = {10.1029/JA083iA04p01611},
  url     = {https://agupubs.onlinelibrary.wiley.com/doi/10.1029/JA083iA04p01611}
}

@book{Banks1973Aeronomy,
  author    = {Banks, Peter M. and Kockarts, Gaston},
  title     = {Aeronomy},
  publisher = {Academic Press},
  address   = {New York},
  year      = {1973},
  url       = {https://shop.elsevier.com/books/aeronomy/banks/978-0-12-077801-0}
}

@article{Shinohara2007Wireless,
  author  = {Shinohara, Naoki},
  title   = {Wireless Power Transmission from Space Solar Power Satellite/Station ({SPS})},
  journal = {The Journal of The Institute of Electrical Engineers of Japan},
  volume  = {129},
  number  = {7},
  pages   = {426--429},
  year    = {2009},
  doi     = {10.1541/ieejjournal.129.426},
  url     = {https://doi.org/10.1541/ieejjournal.129.426}
}

@article{Bilitza2017,
  title={International Reference Ionosphere 2016: From ionospheric climate to real-time weather predictions},
  author={Bilitza, Dieter and Altadill, David and Truhlik, Vladimir and Shubin, Sergey and Galkin, Ivan and Reinisch, Bodo and Huang, Xueqin},
  journal={Space Weather},
  volume={15},
  number={2},
  pages={418--429},
  year={2017},
  doi={10.1002/2016SW001593},
  url={https://agupubs.onlinelibrary.wiley.com/doi/10.1002/2016SW001593}
}

@article{Picone2002,
  title={NRLMSISE-00 empirical model of the atmosphere: Statistical comparisons and scientific issues},
  author={Picone, J. M. and Hedin, A. E. and Drob, D. P. and Aikin, A. C.},
  journal={Journal of Geophysical Research: Space Physics},
  volume={107},
  number={A12},
  pages={1468},
  year={2002},
  doi={10.1029/2002JA009430},
  url={https://doi.org/10.1029/2002JA009430}
}

@book{Gurevich1978,
  title={Nonlinear Phenomena in the Ionosphere},
  author={Gurevich, Aleksandr Viktorovich},
  volume={10},
  year={1978},
  publisher={Springer Science \& Business Media},
  note={Focusing on the ponderomotive redistribution of plasma density.},
  doi={10.1007/978-3-642-87649-3},
  url={https://link.springer.com/book/10.1007/978-3-642-87649-3}
}

@book{Budden1985,
  author    = {Budden, Kenneth George},
  title     = {The Propagation of Radio Waves: The Theory of Radio Waves of Low Power in the Ionosphere and Magnetosphere},
  publisher = {Cambridge University Press},
  address   = {Cambridge, UK},
  year      = {1985},
  doi       = {10.1017/CBO9780511564321},
  url       = {https://doi.org/10.1017/CBO9780511564321}
}

@article{Shinohara2013Rectennas,
  author  = {Shinohara, Naoki},
  title   = {Rectennas for microwave power transmission},
  journal = {IEICE Electronics Express},
  volume  = {10},
  number  = {21},
  pages   = {20132009},
  year    = {2013},
  doi     = {10.1587/elex.10.20132009},
  url     = {https://www.jstage.jst.go.jp/article/elex/10/21/10_10.20132009/_article}
}

@article{Brown1984RadioPower,
  author  = {Brown, William C.},
  title   = {The History of Power Transmission by Radio Waves},
  journal = {IEEE Transactions on Microwave Theory and Techniques},
  volume  = {32},
  number  = {9},
  pages   = {1230--1242},
  year    = {1984},
  doi     = {10.1109/TMTT.1984.1132833},
  url     = {https://doi.org/10.1109/TMTT.1984.1132833}
}

@article{Matsumoto1982SPS,
  author  = {Matsumoto, Hiroshi},
  title   = {Numerical estimation of {SPS} microwave impact on ionospheric environment},
  journal = {Acta Astronautica},
  volume  = {9},
  number  = {8},
  pages   = {493--497},
  year    = {1982},
  doi     = {10.1016/0094-5765(82)90095-9},
  url     = {https://doi.org/10.1016/0094-5765(82)90095-9}
}

@article{Matsumoto1995Nonlinear,
  author  = {Matsumoto, Hiroshi and Hashino, Yoshitaka and Yashiro, Hiroyuki and Shinohara, Naoki and Omura, Yoshiharu},
  title   = {Computer simulation on nonlinear interaction of intense microwave with space plasmas},
  journal = {Electronics and Communications in Japan (Part III: Fundamental Electronic Science)},
  volume  = {78},
  number  = {11},
  pages   = {89--103},
  year    = {1995},
  doi     = {10.1002/ecjc.4430781109},
  url     = {https://doi.org/10.1002/ecjc.4430781109}
}

@article{Kaya1986MINIX,
  author  = {Kaya, Nobuyuki and Matsumoto, Hiroshi and Miyatake, Shinichiro and others},
  title   = {Nonlinear interaction of strong microwave beam with the ionosphere: {MINIX} rocket experiment},
  journal = {Space Solar Power Review},
  volume  = {6},
  pages   = {181--186},
  year    = {1986}
}

@article{Nagatomo1986MINIX,
  author  = {Nagatomo, Makoto and Kaya, Nobuyuki and Matsumoto, Hiroshi},
  title   = {Engineering aspect of the microwave ionosphere nonlinear interaction experiment ({MINIX}) with a sounding rocket},
  journal = {Acta Astronautica},
  volume  = {13},
  number  = {1},
  pages   = {23--29},
  year    = {1986},
  doi     = {10.1016/0094-5765(86)90004-4},
  url     = {https://doi.org/10.1016/0094-5765(86)90004-4}
}

@article{Perkins1981SelfFocusing,
  author  = {Perkins, F. W. and Goldman, M. V.},
  title   = {Self-focusing of radio waves in an underdense ionosphere},
  journal = {Journal of Geophysical Research: Space Physics},
  volume  = {86},
  number  = {A2},
  pages   = {600--608},
  year    = {1981},
  doi     = {10.1029/JA086iA02p00600},
  url     = {https://doi.org/10.1029/JA086iA02p00600}
}

@inproceedings{Omura2005SSPS,
  author    = {Usui, H. and Matsumoto, H. and Omura, Y.},
  title     = {Possible influences of {SSPS} on the space plasma environment},
  booktitle = {Proceedings of the 2nd International Conference on Recent Advances in Space Technologies, 2005. {RAST} 2005},
  pages     = {34--38},
  year      = {2005},
  publisher = {IEEE},
  doi       = {10.1109/RAST.2005.1512530},
  url       = {https://doi.org/10.1109/RAST.2005.1512530}
}

@article{Bonzanini2023MLLTP,
  author  = {Bonzanini, Angelo D. and Shao, Ketong and Graves, David B. and Hamaguchi, Satoshi and Mesbah, Ali},
  title   = {Foundations of machine learning for low-temperature plasmas: Methods and case studies},
  journal = {Plasma Sources Science and Technology},
  volume  = {32},
  number  = {2},
  pages   = {024003},
  year    = {2023},
  doi     = {10.1088/1361-6595/acb28c},
  url     = {https://doi.org/10.1088/1361-6595/acb28c}
}

@article{Zhong2022LTPPINN,
  author  = {Zhong, Linlin and Wu, Bingyu and Wang, Yifan},
  title   = {Low-temperature plasma simulation based on physics-informed neural networks: Frameworks and preliminary applications},
  journal = {Physics of Fluids},
  volume  = {34},
  number  = {8},
  pages   = {087116},
  year    = {2022},
  doi     = {10.1063/5.0106506},
  url     = {https://doi.org/10.1063/5.0106506}
}

@article{Glaser1968Power,
  author  = {Glaser, Peter E.},
  title   = {Power from the Sun: Its Future},
  journal = {Science},
  volume  = {162},
  number  = {3856},
  pages   = {857--861},
  year    = {1968},
  doi     = {10.1126/science.162.3856.857},
  url     = {https://doi.org/10.1126/science.162.3856.857}
}

@techreport{DOENASA1978SPS,
  author      = {{U.S. Department of Energy} and {National Aeronautics and Space Administration}},
  title       = {Satellite Power System Concept Development and Evaluation Program: Reference System Report},
  institution = {U.S. Department of Energy and NASA},
  year        = {1978},
  number      = {DOE/ER-0023},
  address     = {Washington, DC},
  url         = {https://ntrs.nasa.gov/citations/19790014479}
}

@inproceedings{Mankins2012SPSAlpha,
  author    = {Mankins, John C. and Kaya, Nobuyuki and Vasile, Massimiliano},
  title     = {{SPS-ALPHA}: The First Practical Solar Power Satellite via Arbitrarily Large Phased Array (A 2011--2012 {NIAC} Project)},
  booktitle = {10th International Energy Conversion Engineering Conference},
  year      = {2012},
  publisher = {American Institute of Aeronautics and Astronautics},
  doi       = {10.2514/6.2012-3978},
  url       = {https://doi.org/10.2514/6.2012-3978}
}

@book{Hasegawa1975Plasma,
  author    = {Hasegawa, Akira},
  title     = {Plasma Instabilities and Nonlinear Effects},
  publisher = {Springer},
  address   = {Berlin},
  year      = {1975},
  doi       = {10.1007/978-3-642-65980-5},
  url       = {https://doi.org/10.1007/978-3-642-65980-5}
}

@book{Akhiezer1975Plasma,
  author    = {Akhiezer, A. I. and Akhiezer, I. A. and Polovin, R. V. and Sitenko, A. G. and Stepanov, K. N.},
  title     = {Plasma Electrodynamics, Volume 2: Non-Linear Theory and Fluctuations},
  publisher = {Pergamon Press},
  address   = {Oxford},
  year      = {1975}
}

@article{Matsumoto1986Cyclotron,
  author  = {Matsumoto, Hiroshi and Kimura, I.},
  title   = {Nonlinear excitation of electron cyclotron waves by a monochromatic strong microwave: Computer simulation analysis of the {MINIX} results},
  journal = {Space Power},
  volume  = {6},
  pages   = {187--191},
  year    = {1986}
}

@article{Shinohara1995SelfFocusing,
  author  = {Shinohara, Naoki and Shklyar, David R. and Matsumoto, Hiroshi},
  title   = {Numerical analysis of self-focusing effect caused by inhomogeneity of microwave energy density in ionosphere},
  journal = {Electronics and Communications in Japan (Part I: Communications)},
  volume  = {79},
  number  = {9},
  pages   = {82--93},
  year    = {1996},
  doi     = {10.1002/ecja.4410790910},
  url     = {https://doi.org/10.1002/ecja.4410790910}
}

@mastersthesis{Nakamoto2007Cavitation,
  author = {Nakamoto, N.},
  title  = {Study on Space Plasma Disturbances Induced by Spatial Intensity Gradients of Large-Amplitude Electromagnetic Wave Beams},
  school = {Department of Electrical Engineering, Graduate School of Engineering, Kyoto University},
  year   = {2007},
  month  = feb
}

@book{Monk2003MaxwellFEM,
  author    = {Monk, Peter},
  title     = {Finite Element Methods for Maxwell's Equations},
  publisher = {Oxford University Press},
  address   = {Oxford},
  year      = {2003},
  doi       = {10.1093/acprof:oso/9780198508885.001.0001},
  url       = {https://doi.org/10.1093/acprof:oso/9780198508885.001.0001}
}

@book{Jin2014FEMElectromagnetics,
  author    = {Jin, Jian-Ming},
  title     = {The Finite Element Method in Electromagnetics},
  edition   = {3},
  publisher = {Wiley-IEEE Press},
  address   = {Hoboken, NJ},
  year      = {2014},
  isbn      = {978-1-118-57136-1}
}

@article{Berenger1994PML,
  author  = {Berenger, Jean-Pierre},
  title   = {A Perfectly Matched Layer for the Absorption of Electromagnetic Waves},
  journal = {Journal of Computational Physics},
  volume  = {114},
  number  = {2},
  pages   = {185--200},
  year    = {1994},
  doi     = {10.1006/jcph.1994.1159},
  url     = {https://doi.org/10.1006/jcph.1994.1159}
}

@article{Fritsch1980PCHIP,
  author  = {Fritsch, F. N. and Carlson, R. E.},
  title   = {Monotone Piecewise Cubic Interpolation},
  journal = {SIAM Journal on Numerical Analysis},
  volume  = {17},
  number  = {2},
  pages   = {238--246},
  year    = {1980},
  doi     = {10.1137/0717021},
  url     = {https://doi.org/10.1137/0717021}
}

@article{Bilitza2022IRIReview,
  author  = {Bilitza, Dieter and Pezzopane, Michael and Truhlik, Vladimir and Altadill, David and Reinisch, Bodo W. and Pignalberi, Alessio},
  title   = {The International Reference Ionosphere Model: A Review and Description of an Ionospheric Benchmark},
  journal = {Reviews of Geophysics},
  volume  = {60},
  number  = {4},
  pages   = {e2022RG000792},
  year    = {2022},
  doi     = {10.1029/2022RG000792},
  url     = {https://doi.org/10.1029/2022RG000792}
}

@article{Emmert2021NRLMSIS20,
  author  = {Emmert, J. T. and Drob, D. P. and Picone, J. M. and Siskind, D. E. and Jones, M. and Mlynczak, M. G. and Bernath, P. F. and Chu, X. and Doornbos, E. and Funke, B. and Goncharenko, L. P. and Hervig, M. E. and Schwartz, M. J. and Sheese, P. E. and Vargas, F. and Williams, B. P. and Yuan, T.},
  title   = {{NRLMSIS} 2.0: A Whole-Atmosphere Empirical Model of Temperature and Neutral Species Densities},
  journal = {Earth and Space Science},
  volume  = {8},
  number  = {3},
  pages   = {e2020EA001321},
  year    = {2021},
  doi     = {10.1029/2020EA001321},
  url     = {https://doi.org/10.1029/2020EA001321}
}

@article{Raissi2019PINN,
  author  = {Raissi, Maziar and Perdikaris, Paris and Karniadakis, George Em},
  title   = {Physics-informed Neural Networks: A Deep Learning Framework for Solving Forward and Inverse Problems Involving Nonlinear Partial Differential Equations},
  journal = {Journal of Computational Physics},
  volume  = {378},
  pages   = {686--707},
  year    = {2019},
  doi     = {10.1016/j.jcp.2018.10.045},
  url     = {https://doi.org/10.1016/j.jcp.2018.10.045}
}

@article{Sasaki2013MPT,
  author  = {Sasaki, Susumu and Tanaka, Koji and Maki, Ken-ichiro},
  title   = {Microwave Power Transmission Technologies for Solar Power Satellites},
  journal = {Proceedings of the IEEE},
  volume  = {101},
  number  = {6},
  pages   = {1438--1447},
  year    = {2013},
  doi     = {10.1109/JPROC.2013.2246851},
  url     = {https://doi.org/10.1109/JPROC.2013.2246851}
}

@article{Kaya1981SpaceChamber,
  author  = {Kaya, Nobuyuki and Matsumoto, Hiroshi},
  title   = {Space Chamber Experiments of Ohmic Heating by High Power Microwave from the Solar Power Satellite},
  journal = {Geophysical Research Letters},
  volume  = {8},
  number  = {12},
  pages   = {1289--1292},
  year    = {1981},
  doi     = {10.1029/GL008i012p01289},
  url     = {https://doi.org/10.1029/GL008i012p01289}
}

@article{Perkins1974Parametric,
  author  = {Perkins, F. W. and Oberman, C. and Valeo, E. J.},
  title   = {Parametric Instabilities and Ionospheric Modification},
  journal = {Journal of Geophysical Research},
  volume  = {79},
  number  = {10},
  pages   = {1478--1496},
  year    = {1974},
  doi     = {10.1029/JA079i010p01478},
  url     = {https://doi.org/10.1029/JA079i010p01478}
}

@techreport{Duncan1977IonosphereMicrowave,
  author      = {Duncan, Lewis M. and Gordon, Wm. E.},
  title       = {Ionosphere/Microwave Beam Interaction Study},
  institution = {William Marsh Rice University},
  type        = {Final Report},
  number      = {NASA-CR-151517},
  address     = {Houston, TX},
  year        = {1977},
  url         = {https://ntrs.nasa.gov/citations/19770024381}
}

@article{Thome1974Striations,
  author  = {Thome, G. D. and Perkins, F. W.},
  title   = {Production of Ionospheric Striations by Self-Focusing of Intense Radio Waves},
  journal = {Physical Review Letters},
  volume  = {32},
  number  = {22},
  pages   = {1238--1241},
  year    = {1974},
  doi     = {10.1103/PhysRevLett.32.1238},
  url     = {https://doi.org/10.1103/PhysRevLett.32.1238}
}

@article{Malaviya2022SSPSReview,
  author  = {Malaviya, Preyansh and Sarvaiya, Vishwadeep and Shah, Abhishek and Thakkar, Drupad and Shah, Manan},
  title   = {A Comprehensive Review on Space Solar Power Satellite: An Idiosyncratic Approach},
  journal = {Environmental Science and Pollution Research},
  volume  = {29},
  pages   = {42476--42492},
  year    = {2022},
  doi     = {10.1007/s11356-022-19560-w},
  url     = {https://doi.org/10.1007/s11356-022-19560-w}
}

\end{document}